\documentclass[twocolumn]{aastex631}
\usepackage{float}
\usepackage{booktabs}
\usepackage{threeparttable} 
\usepackage{multirow} 
\usepackage{graphicx}
\usepackage{amsmath} 
\usepackage{amssymb}
\usepackage{subfigure}
\usepackage{hyperref} 
\bibpunct{(}{)}{;}{a}{}{,}  
\makeatletter
\renewcommand{\@thesubfigure}{\hskip\subfiglabelskip}
\makeatother
\makeatletter

\graphicspath{{./}{Fig/}}

\received{October 28, 2023}
\revised{February 13, 2024}
\submitjournal{ApJ}
\accepted{February 29, 2024}

\shorttitle{Variation study for ON 231}
\shortauthors{Wang $\&$ Jiang}

\begin{document}
\title{Revealing the variation mechanism of ON 231 via the two-components shock-in-jet model}
\author[0009-0002-5955-4932]{Chi-Zhuo Wang}
\affiliation{Shandong Provincial Key Laboratory of Optical Astronomy and Solar-Terrestrial Environment,\\
Institute of Space Sciences, Shandong University, Weihai, 264209, China}
\author[0000-0003-2679-0445]{Yun-Guo Jiang}
\affiliation{Shandong Provincial Key Laboratory of Optical Astronomy and Solar-Terrestrial Environment,\\
Institute of Space Sciences, Shandong University, Weihai, 264209, China}

\begin{abstract}
The variation mechanism of blazars is a long-standing unresolved problem. In this work, we present a scenario to explain diverse variation phenomena for ON 231, where the jet emissions are composed of the flaring and the less variable components (most probably from the post-flaring blobs), and the variation is dominated by shock-in-jet instead of the Doppler effect. We perform correlation analysis for the multiwavelength light curves and find no significant correlations. For optical band, ON 231 exhibits a harder when brighter (HWB) trend, and the trend seems to shift at different periods. Correspondingly, the correlation between polarization degree and flux exhibits a V-shaped behavior, and a similar translation relation during different periods is also found. These phenomena could be understood via the superposition of the flaring component and slowly varying background component. We also find that the slopes of HWB trend become smaller at higher flux levels, which indicates the energy-dependent acceleration processes of the radiative particles. For X-ray, we discover a trend transition from HWB to softer when brighter (SWB) to HWB. We consider that the X-ray emission is composed of both the synchrotron tail and the Synchrotron Self-Compton components, which could be described by two log-parabolic functions. By varying the peak frequency, we reproduce the observed trend transition in a quantitative manner. For $\gamma$-ray, we find the SWB trend, which could be explained naturally if a very-high-energy $\gamma$-ray background component exists.  Our study elucidates the variation mechanism of intermediate synchrotron-peaked BL Lac objects.

\end{abstract}


\section{Introduction} \label{sec:intro}
Blazars are a subclass of active galactic nuclei (AGN) with relativistic jets pointing to our line of sight \citep{1995PASP..107..803U}. They display violent flux, polarization, and spectral variability at all wavelengths. According to the optical emission line features, they can be divided into two subclasses, including the flat-spectrum radio quasars (FSRQs) and BL Lacaerte objects (BL Lacs). The spectral energy distributions (SEDs) of blazars exhibit two bumps. It is generally accepted that the low energy bump is dominated by synchrotron radiation, while the high energy bump is produced by the inverse Compton scattering process \citep{2009ApJ...704...38S,2013ApJ...768...54B,2017SSRv..207....5R}. The radiation could be from multiple background components such as the accretion disk, dusty torus, and broad line region \citep{2019A&A...627A..72G}, and the variation could be affected by many factors including geometric effects, magnetic reconnection, and shock. Thus, the variation mechanism of blazars at various timescales is complex and under intensive debate. A comprehensive study of various aspects of the variation, including flux, spectrum, polarization degree (PD), and polarization angle (PA) can help us constrain the variation mechanism and understand the essential nature of blazars.

ON 231 (also known as W Comae, W Com, 1219+285) is a violently variable BL Lac object in Coma Berenices. It was discovered as a variable star by \citet{1916AN....202..415W} and identified as the counterpart of a radio source by \citet{1971Natur.231..515B}. An accurate redshift (z = 0.102) was determined by \citet{2006A&A...445..441N}. ON 231 was classified as an intermediate synchrotron-peaked BL Lac (IBL) object, since its low energy bump of SED peak at optical band \citep{2006A&A...445..441N,2010ApJ...716...30A}. The very long baseline interferometry (VLBI) observations of ON 231 revealed its structure with a core and a multi-component jet elongated in Position Angle $\sim110^{\circ}$ \citep{1985ApJ...292..614W,1992ApJ...388...40G,1994ApJ...435..140G,1996MNRAS.283..759G,2001A&A...374..435M}. The historical optical light curves of ON 231 exhibited variability at various timescales ranging from a few hours to several years \citep{1992ApJS...80..683X,1995PASP..107..863S,2012MNRAS.425.3002G}. Since the beginning of the last century, the source has shown several optical outbursts in 1940, 1953, and 1968 \citep{1998A&AS..130..109T}. After that, the source started brightening slowly and reached its most luminous state in 1988 with $R$ = 12.2 mag \citep{1999A&A...342L..49M}. After the exceptional optical outburst occurred in 1998, a large bend in the jet at about 10 mas from the core was observed by the European VLBI Network plus MERLIN at 1.6 GHz and 5 GHz \citep{2001A&A...374..435M}. ON 231 was discovered as an extragalactic very-high-energy $\gamma$-ray emitter by VERITAS in 2008 March, and it entered into a second outburst state in June of the same year with the $\gamma$-ray flux about three times brighter than the previous one detected in 2008 March \citep{2008ApJ...684L..73A,2009ApJ...707..612A}. \citet{2014ApJ...794...54S} reported a large rotation of PA from $78^{\circ}$ to $315^{\circ}$ ($\Delta\theta \sim 237^{\circ}$) that coincided in time with the second $\gamma$-ray outburst. Combined with the Stokes parameters, they inferred that two optically thin synchrotron components contributed to the polarized flux. 
\citet{2022MNRAS.517.3236R} studied the correlation between optical flux and polarization during ten cycles observed by the Steward Observatory. The source exhibited significant positive correlations only in two cycles. At short-term timescales, they also found two significant correlations (one is positive and the other is negative).
Additionally, many studies show that ON 231 exhibits evident bluer when brighter (hereafter BWB) behaviors, just as most BL Lacs \citep{2018ApJS..237...30M,2013MNRAS.429.2773C,1998A&AS..130..109T,2008PASJ...60..145Z}. \citet{2023MNRAS.519.5263Z} found the source became bluer and then gradually stabilized when it brightened, which was briefly named the bluer-stable-when-brighter (BSWB). They inferred that the optical emission was mainly composed of two stable-color components. Generally speaking, ON 231 exhibits complex variation behaviors at various timescales, and there is still a lack of a convincing theoretical framework to elucidate the variations.

In this work, we make a comprehensive study on various variation phenomena for this target. We also propose a theoretical framework, which could consistently explain these phenomena and reveal the variation mechanism in an analytical manner. This paper is organized as follows. In Section \ref{sec:data}, the multiwavelength data are collected and reduced. In Section \ref{sec:Var}, the multiwavelength light curves from $\gamma$-ray to radio are presented in Section \ref{subsec:LC}. The results of the local cross-correlation analysis are presented in Section \ref{subsec:CC}. In Section \ref{subsec:Spectral}, we discuss various variation phenomena, including the $\gamma$-ray photon index (hereafter PI), X-ray PI, optical spectral index, color index (hereafter CI), and broadband SED. A theoretical framework is presented to describe these behaviors and reveal the variation mechanism. In Section \ref{subsec:PD}, we study the correlation between PD with flux and the rotation behavior of PA on $qu$-plane. The above behaviors at short-term timescales are also investigated in Section \ref{subsec:Sta}. Finally, we review the complex variation behavior at various timescales and present the conclusion in Section \ref{sec:conclusion}.

\section{DATA COLLECTION} \label{sec:data}
In this work, we collect multiwavelength data from public data archives, including the $\gamma$-ray data from the Fermi LAT Light Curve Repository (LCR; \citealt{2023ApJS..265...31A}), the X-ray data from Swift \citep{2004ApJ...611.1005G}, the optical data from the All-Sky Automated Survey for Supernovae (ASAS; \citealt{2014ApJ...788...48S}), the optical photometry, spectroscopy, and polarization data from Steward Observatory (SO; \citealt{2009arXiv0912.3621S}), and the radio 15 GHz data from the Owen Valley Radio Observatory (OVRO; \citealt{2011ApJS..194...29R}).

\begin{figure*}
	\centering
	\includegraphics[width=1.55\columnwidth]
 {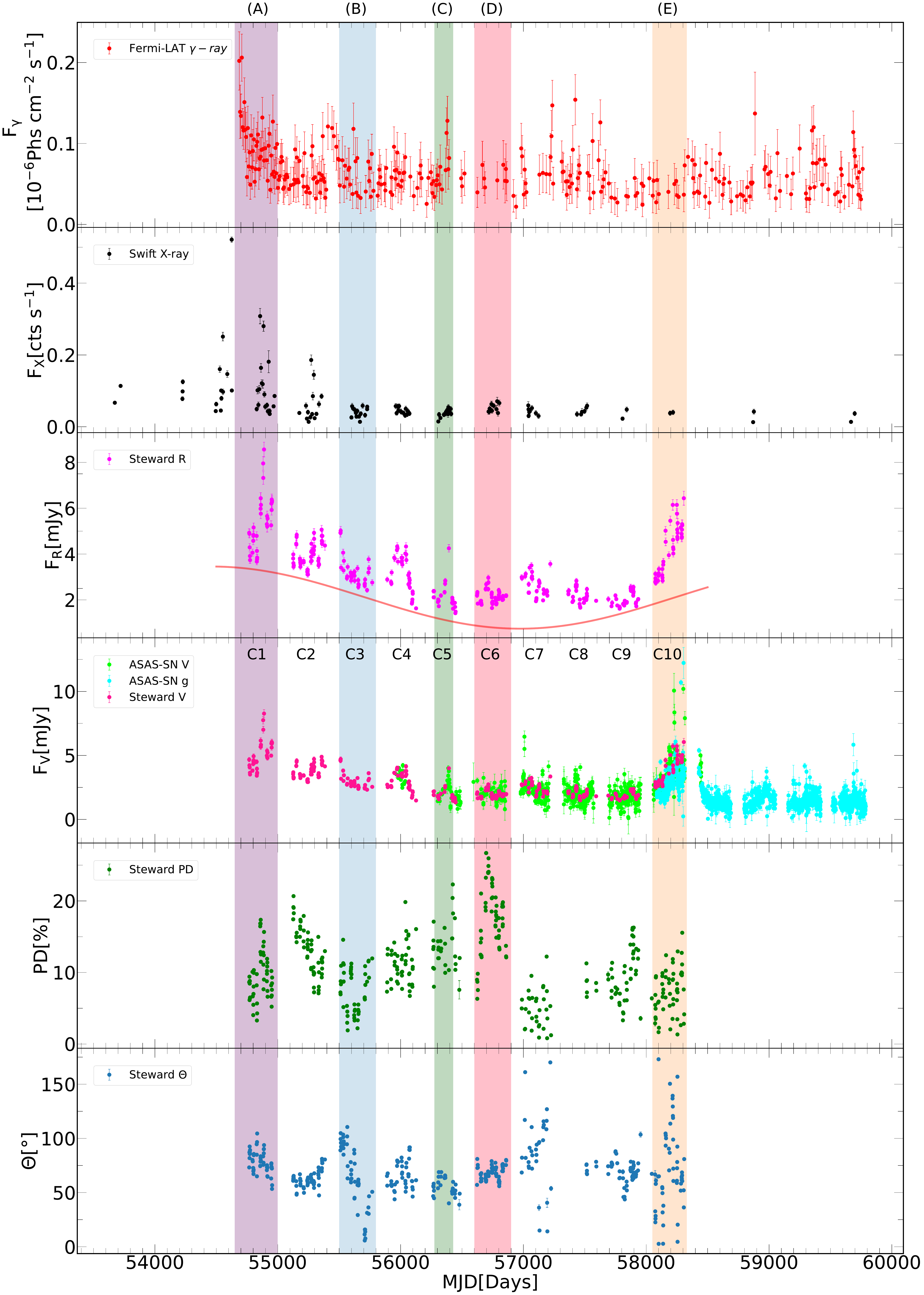}
	\caption{From top to bottom, panels exhibit the light curves of $\gamma$-ray, X-ray, optical $V$-band, optical $R$-band, PD, PA, and radio 15GHz, respectively. Unfortunately, we are unable to show radio data obtained from the OVRO directly in the publication due to some data usage policy. C1$\sim$ C10 represent the periods divided by the SO sampling interval. Different background colors also represent different periods. The purple zone is named period $(A)$, the blue zone is named period $(B)$, the green zone is named period $(C)$, the red zone is named period $(D)$, and the orange zone is named period $(E)$.}
\label{lc}
\end{figure*}

{\textbf{\emph{Fermi} LAT data:}} We collect nearly 14 years (from 2008 August 8 to 2022 June 8) $\gamma$-ray data of ON 231 from the LCR \citep{2023ApJS..265...31A}\footnote{\url{https://fermi.gsfc.nasa.gov/ssc/}}. The LCR is a database of multi-cadence flux calibrated light curves for over 1500 sources deemed variable in the 10-year Fermi LAT point source (4FGL-DR2) catalog \citep{2020arXiv200511208B}. The photons are selected from a $12^{\circ}$ radius energy-independent region of interest (ROI) centered on the target. The photon energy range is from $0.1$ to $100$ GeV. The data analysis is performed with the standard Fermi LAT science tools version v11r5p3 using a maximum likelihood analysis \citep{2009ApJS..183...46A}. The instrument response functions (IRFs), Galactic diffuse background, and isotropic $\gamma$-ray background files are {P8R2\_SOURCE\_V6}, {gll\_iem\_v07.fits}, and iso {P8R3\_SOURCE\_V3\_v1.txt}, respectively. We choose one week as the time bin and use the test statistics (TS) to filter the data. Finally, we obtain $723$ data points with TS \textgreater 10. We also obtain the $\gamma$-ray PI fitted by the power-law model. The formula is given in $dN/dE=N_{0}(E/E_{0})^{-\Gamma}$, here, the PI is defined as $-\Gamma$.

\textbf{X-ray data:} We collect the X-ray data from the XRT \citep{2005SSRv..120..165B} mounted on the Swift satellite \citep{2004ApJ...611.1005G}, which operates in the energy range of $0.3-10$ keV. All the observations are performed in Photon Counting (PC) mode. We analyze the data with the \textit{xrtpipeline} task, utilizing the latest CALDB and response files provided in HEASOFT version 6.32. The source spectra are extracted from a circular area with a radius of $50^{\prime\prime}$, positioned at the center of the source, whereas, the background spectra are extracted from an annulus region with an inner radius of $70^{\prime\prime}$ and an outer radius of $130^{\prime\prime}$. We perform the X-ray spectral analysis within \textit{XSPEC} \citep{1996ASPC..101...17A} and employ an absorbed simple power law model with a Galactic neutral hydrogen column density $N_{H}=1.95 \times 10^{20}$ cm$^{-2}$ \citep{2016A&A...594A.116H,2005A&A...440..775K,1990ARA&A..28..215D}. Finally, we obtain the X-ray PI and flux in different energy bands, including $0.3-10$ keV, $0.3-1.5$ keV, and $2.5-10$ keV. Here, we define the soft band as $0.3-1.5$ keV, and the hard band as $2.5-10$ keV. The hardness ratio (HR) is calculated by $HR=H/S$, where $H$ and $S$ are the flux in the defined hard and soft bands, respectively.

\textbf{Optical data:} The photometry and polarization data are mainly taken from the SO \citep{2009arXiv0912.3621S}\footnote{\url{http://james.as.arizona.edu/~psmith/Fermi/}} which is an important part of the Ground-based Observational Support of the Fermi Gamma-ray Space Telescope. The duration of $V$-band, $R$-band, and polarization data observed by the 2.3 m Bok Telescope and the 1.54 m Kuiper Telescope is from 2008 October to 2018 July. All these data obtained by the SPOL CCD Imaging or Spectropolarimeter have been calibrated. We also retrieve the $V$-band and $g$-band datas from the ASAS \citep{2014ApJ...788...48S}\footnote{\url{https://asas-sn.osu.edu/}}. For the above magnitude data, we convert them into flux data ulteriorly.

Additionally, we download the optical spectroscopy data from the SO and correct the Galactic interstellar extinction and reddening of all spectra by the procedure given in \citet{1989ApJ...345..245C}. We chose a line free window at 5100 Å to extract the density flux $F_{\lambda}$ and translate it to $F_{\nu}$ [mJy]. To obtain the optical spectral index $\alpha_{o}$, we represent the data in the $\log{\nu}$ - $\log{\nu}F_{\nu}$ plane and manually select six frequencies from the time integrated spectrum. These six frequencies are in the line (either emission or absorption) free regime, and the fluxes at them are of local minimum. We take the slope of the linear fitting result as $\alpha_{o}$ for each spectrum. We also abandon the air absorption window at 6800 - 7500 Å for analysis. The spectral behavior for this target is exhibited in the plot of $\alpha_{o}$ versus $\log F_{\nu}$ in Figure \ref{SI}.

\textbf{Radio 15 GHz data:} The radio 15 GHz data from 2008 January 28th to 2020 January 25th are collected from the OVRO 40 m monitoring program \citep{2011ApJS..194...29R}\footnote{\url{http://www.astro.caltech.edu/ovroblazars/}}. The data point sampling of this light curve is relatively continuous. There are 600 data points in total with at least one data point per week on average. However, we couldn't display them directly in Figure \ref{lc} due to the data usage policy reasons. The data may be available on request to the OVRO 40 m collaboration.

\section{VARIATION ANALYSIS} \label{sec:Var}

\begin{figure*}[t!]
	\centering 
	\subfigure[$(a)$]{
		\includegraphics[width=5.5cm]{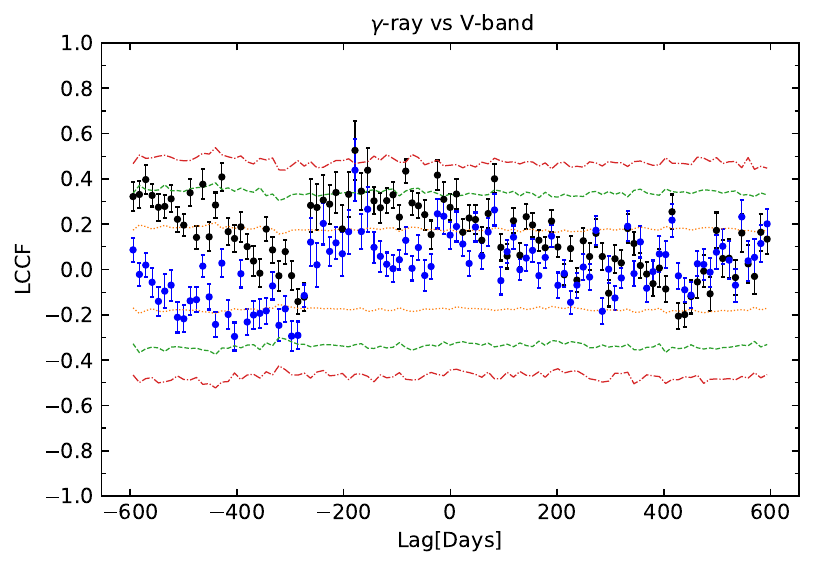}}
	\quad
	\subfigure[$(b)$]{
		\includegraphics[width=5.5cm]{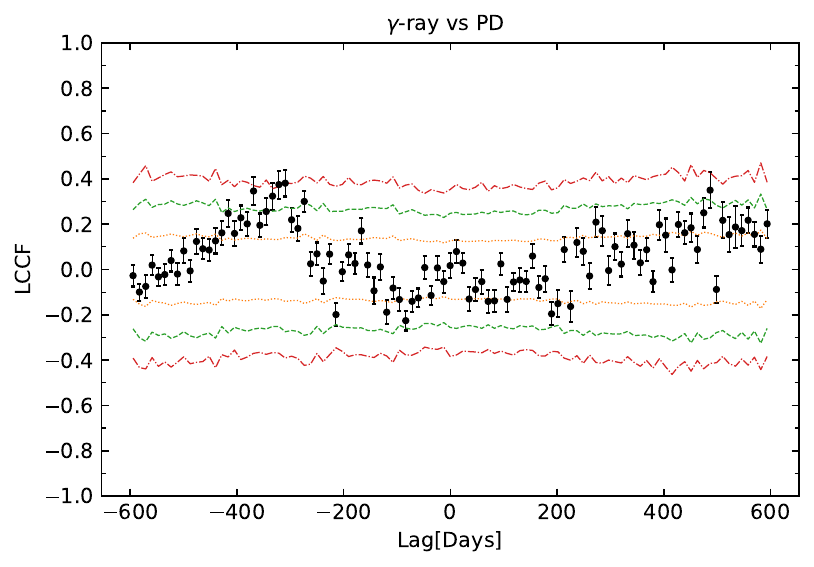}}
	\quad
	\subfigure[$(c)$]{
		\includegraphics[width=5.5cm]{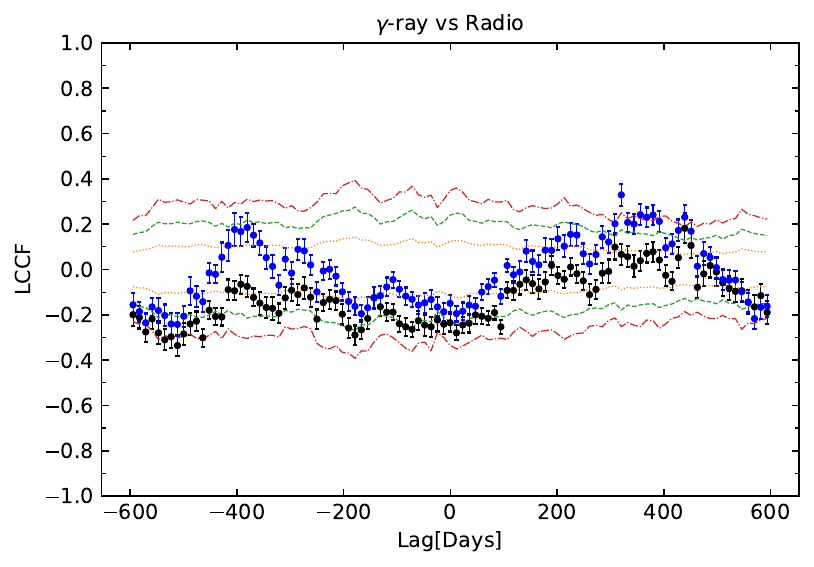}}
	\quad
	
	\subfigure[$(d)$]{
		\includegraphics[width=5.5cm]{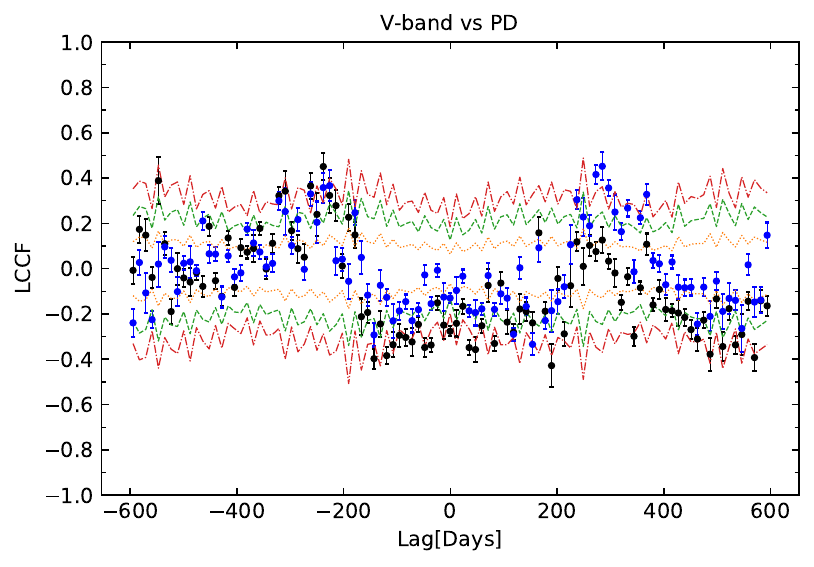}}
	\quad
	\subfigure[$(e)$]{
		\includegraphics[width=5.5cm]{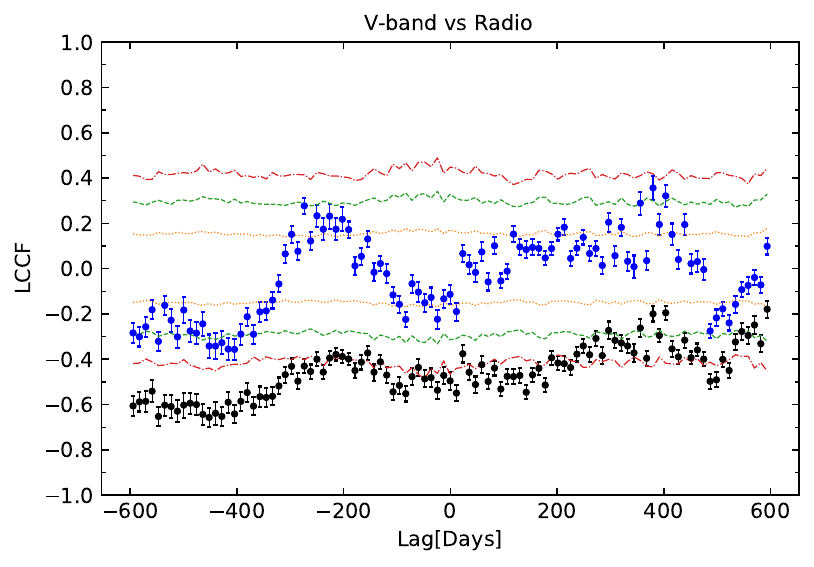}}
	\quad
	\subfigure[$(f)$]{
		\includegraphics[width=5.5cm]{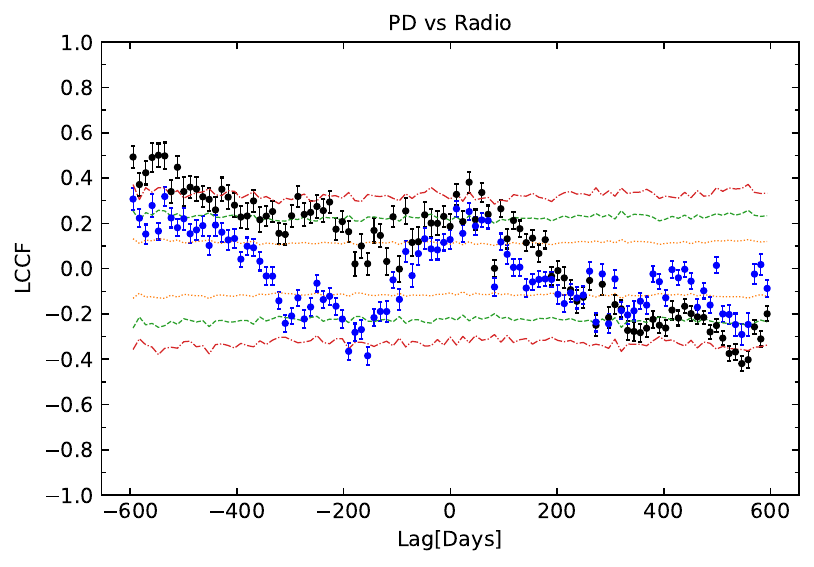}}
	\caption{The LCCF analysis results among light curves of different wavebands are plotted. The significance levels of 1$\sigma $, 2$\sigma $, and 3$\sigma $ are indicated by the orange, green, and red lines, respectively. The blue dots in some panels represent the LCCF results based on the corrected radio and optical light curves.}
	\label{LCCF}
\end{figure*}

\subsection{Character of Light Curves} \label{subsec:LC}
The multiwavelength light curves of ON 231 from MJD 53550 to 59800 are plotted in Figure \ref{lc}. The light curve of $\gamma$-ray is shown on the top panel of Figure \ref{lc}. Combined with previous research, it can be seen that the $\gamma$-ray flux gradually drops down to a low state ($ F_{\gamma}$ \textless $10^{-7} $ Phs cm$ ^{-2} $ s$ ^{-1} $) after a series of very high-energy flares during 2008 \citep{2008ApJ...684L..73A,2009ApJ...707..612A}. At the same time, the sparse X-ray data exhibits similar behavior, i.e., the source gradually becomes quiet after the outburst occurred in MJD 54625. On the third and fourth panels of Figure \ref{lc}, the optical observation from the SO shows that the flux of $R$-band ($F_{R}$) varies from $1.4$ mJy to $8.6$ mJy while the flux of $V$-band ($F_{V}$) varies from $1.3$ mJy to $8.3$ mJy. It has a minimum brightness with $F_{R}=1.42$ mJy ($F_{V}=1.41$ mJy) around February 26th, 2009, and a maximum brightness with $F_{R}=8.56$ mJy ($F_{V}=8.27$ mJy) around June 5, 2013. The maximum flux changes are $\Delta F_{V} \sim6.86$ mJy and $\Delta F_{R} \sim7.14$ mJy, which occurs on a timescale of $\sim1560$ days ($\sim4.3$ yr). According to the data from the ASAS, we notice that there is a remarkable flare with the flux of $g$-band up to 12.23 mJy on July 5th, 2018 which is brighter than the last outburst on February 26th, 2009. Throughout the photometric history, ON 231 has exhibited several notable outbursts in 1940, 1953, 1968, 1975, 1987, 1998, 2008 \citep{1979AJ.....84.1658P,1989A&A...220...89X,1992IAUC.5458....2M,1998A&AS..130..109T,2008ApJ...684L..73A}. From this historical record, it is likely that there is an optical periodicity of about ten years. Additionally, it’s worth noting that there are many short-term flares superimposed on long-term trends. Given this, we use the trigonometric function and impulsive function to fit the long-term variation in the optical band and radio band, respectively. We plot the fitted trigonometric function curve with red solid lines on the third panel of Figure \ref{lc} and the fitted impulsive function curve with blue solid lines on the last panel of Figure \ref{lc} (not shown due to the OVRO data policy).

To study the variation behavior at various timescales, we select five periods from the time series based on the optical sampling interval and activity states. In Figure \ref{lc}, period $(A)$ from MJD 54650 to 55000 is marked with purple background, during which there is a giant flare at high luminosity state while period $(C)$ from MJD 56275 to 56428 is marked with green background, during which there is a weak flare at low luminosity state. Period $(D)$ from MJD 56600 to 56900 marked with red background can represent a stable flux state with a mini flare. Period $(B)$ from MJD 55500 to 55800 marked with blue background can represent a decaying part of flare while period $(E)$ from MJD 58050 to 58330 marked with orange background can represent a rising part of flare. Additionally, we divide the optical light curves into ten cycles from C1 to C10 based on the SO sampling interval. In summary, each period represents a different state. We will analyze the variation at various timescales in the following.

\subsection{Cross-Correlation Analysis}\label{subsec:CC}
To our best knowledge, for two separate uneven sampled time series, there are quite a few analysis methods such as the interpolated cross-correlation function (ICCF; \citealt{1987ApJS...65....1G}), the discrete Fourier transform (DFT; \citealt{1989ApJ...343..874S}), and the z-transformed discrete correlation function (ZDCF; \citealt{1997ASSL..218..163A}). Most researchers mainly use the discrete correlation function (DCF; \citealt{1988ApJ...333..646E}) and the localized cross-correlation function (LCCF; \citealt{1999PASP..111.1347W}) to calculate the correlation coefficient between the multiwavelength light curves. Comparing the DCF and LCCF, previous studies suggest that the LCCF is more efficient than the DCF in picking up the physical signal \citep{2014MNRAS.445..437M}. Besides, considering that the LCCF exhibits a constraint on the [-1,1] range of the correlation coefficient, we finally select the LCCF to perform the correlation analysis between light curves at different bands. 

The Monte Carlo (MC) procedure is applied to estimate the significance of the time lag \citep{2019ApJ...884...15S}. For two light curves, we will first choose the better sampling one and simulate 10000 artificial light curves by the method given in \citet{1995A&A...300..707T} (TK95). For ON 231, the spectral slope of the power spectral density (PSD) is computed by the program based on the actual light curves, among which $\beta_{\rm \gamma}=0.75$, $\beta_{optical}=0.21$, $\beta_{radio}=0.66$. During the simulated procedure, we take the time interval of artificial light curves as one day. Afterward, we calculate the LCCFs between the artificial light curves and observed light curves. We obtain the $1\sigma$, $2\sigma$, and $3\sigma$ confidence levels corresponding to the 68.27\%, 95.45\%, and 99.73\% chance probabilities to evaluate the significance. The correlation peak is significant if it exceeds the $3\sigma$ level.

Based on the above methods, we perform correlation analysis on the multiwavelength light curves and the results are shown in Figure \ref{LCCF}. To avoid spurious signals caused by the profile of the light curves, we acquire the detrend light curves of optical and radio bands. Then, we perform the cross-correlation analysis based on the detrend light curves. We find that the LCCF values usually increase and some peaks with physical significance will appear after subtracting the long-term trends. In Figure \ref{LCCF}, a peak at about 400 days appears on panel $(c)$, and a peak at about 250 days appears on panel $(e)$, even though the correlation peaks remain below the $3\sigma$ level. Moreover, a negative correlation exists between PD and optical $V$-band at near-zero time lag. The correlation between X-ray and other bands is too scattered to be shown in Figure \ref{LCCF}. As a whole, we fail to identify any significant ($3\sigma$) correlations between different bands. This indicates that the variation in multiwavelength may be caused by the superposition of emission from multi-components.

\begin{table}
	\begin{center}
		\setlength{\abovecaptionskip}{0.2cm} 
		\setlength{\belowcaptionskip}{-0.2cm}
		\caption{The results of time lag analysis}
		\label{Time lag}
		\centering
        \setlength\tabcolsep{1.2mm}{
		\begin{tabular}{cccc}
			\toprule
			\toprule
			                       & $\tau_{\rm p}$ (days) & $\tau_{\rm c}$ (days) & $\bar{\tau}$ (days) \\
			\midrule
			$\gamma$-ray vs $V$-band &   ${-176.5}^{+161}_{\rm -63}$                   &${-154.2}^{+85.5}_{\rm -50.6}$                       & ${-165.4}^{+123.3}_{\rm -56.8}$                                        \\
		$\gamma$-ray vs Radio  & ${-397.5}^{+46}_{\rm -10}$ &${-378}^{+31.4}_{\rm -17.3}$           & ${-387.8}^{+38.7}_{\rm -13.7}$                    \\
			$V$-band vs Radio        & ${-253.5}^{+49}_{\rm -21}$                      & ${-239.6}^{+27.7}_{\rm -31.8}$                      &                     ${-246.6}^{+38.4}_{\rm -26.4}$                    \\	
			
			\bottomrule
		\end{tabular}}
	\end{center}
\textbf{Notes.} The negative lags indicate that the former leads the latter. $\tau_{\rm p}$ is the lag corresponding to the highest peak of LCCF while $\tau_{\rm c}$ is the centroid lag of the LCCF defined as $\tau_{\rm c}=\sum_{i}\tau_{\rm i}C_{\rm i}$/$\sum_{i}C_{\rm i}$ where $C_{\rm i}$ is the correlation coefficient satisfying $C_{\rm i}$ \textgreater\ $ 0.8$ LCCF $(\tau_{\rm p})$ \citep{2019ApJ...884...15S}. $\bar{\tau}$ is the average of $\tau_{\rm p}$ and $\tau_{\rm c}$.
\end{table}

Based on the $2\sigma$ correlation results, we take the flux randomization and the random subset selection processes into account \citep{1998PASP..110..660P,2012arXiv1207.1459L}, and calculate two kinds of time lag, namely $\tau_{\rm p}$ and $\tau_{\rm c}$. $\tau_{\rm p}$ is the lag corresponding to the highest peak of LCCF while $\tau_{\rm c}$ is the centroid lag of the LCCF defined as $\tau_{\rm c}=\sum_{i}\tau_{\rm i}C_{\rm i}$/$\sum_{i}C_{\rm i}$ where $C_{\rm i}$ is the correlation coefficient satisfying $C_{\rm i}$ \textgreater\ $ 0.8$ LCCF $(\tau_{\rm p})$ \citep{2019ApJ...884...15S}. The time lags are listed in Table \ref{Time lag}. In general, the $\gamma$-ray band leads the $V$-band while the $V$-band leads the radio band. The physics model to explain the time lags between the light curves of different bands is that the photons from different bands are emitted to the outside in different regions of jet \citep{2011MNRAS.415.1631K,2014MNRAS.445..428M}. The distance between different emission regions is calculated according to the formula: $D=\beta_{\rm app}c{\Delta T}/(1+z)\sin{\xi}$, where $\beta_{\rm app}$ is the apparent velocity in the observer frame, $c$ is the light speed, $\Delta T$ is the time lag, $z$ is the redshift, and $\xi$ is the viewing angle between jet axis and observing line of sight. Therefore, results indicate that the optical emitting region may be upstream of the radio emitting region and downstream of the $\gamma$-ray emitting region.

\subsection{Spectral Variability}\label{subsec:Spectral}
Based on the above cross-correlation analysis in Section \ref{subsec:CC}, we can notice that the correlation between different time series is not significant. Thus, the one-zone emission model is not natural to explain the complex variability of the target. We consider the two-component model and try to explain the behavior of the $\gamma$-ray PI, X-ray PI, optical spectral index, CI and broadband SED in the framework of this model. Here, the model is to say that the observed flux contains two components,
\begin{equation}\label{eq:TC}
	{F_{\rm obs}=F_{j}+F_{b}},
\end{equation}
where $F_{j}$ is a flaring component and $F_{b}$ is a less variable background component (most probably from the post-flaring blobs). For most BL Lacs, the shape of bumps in the $\log \nu F_{\nu}$ vs $\log \nu$ plot is characterized by a rather smooth curvature extending through several frequency decades, which can be well described by a log-parabolic function. \citet{2004A&A...413..489M} provides a physical explanation in terms of statistical particle acceleration. In the jet-comoving frame, the log-parabolic model to describe the SED is given in
\begin{gather}\label{eq:LP0} 
	{\log \left(\nu^{\prime} F_{\nu^{\prime}}^{\prime}\right)=\log \left(\nu_{p}^{\prime} F_{\nu_{p }^{\prime}}^{\prime}\right)-b\left[\log \left(\frac{\nu^{\prime}}{\nu_{p}^{\prime}}\right)\right]^{2}},
\end{gather}
where $\nu_{p}^{\prime}$ is the intrinsic peak frequency corresponding to the maximum $\nu_{p}^{\prime} F_{\nu_{p}^{\prime}}$ of the SED and $b$ is the curvature index. Considering the Doppler enhancement, the observed flux can be described as $F_{\nu}= \delta^{m}F_{\nu^{\prime}}^{\prime}$, and the peak frequency in the observing frame is given as $\nu_{p}= \delta \nu_{p}^{\prime}$, where the $\delta$ is the Doppler factor. $m$ takes the value 3. The flux contributed by jet in the observing frame is given in \citet{Jiang2023}
\begin{gather}\label{eq:LP}
	{\log\!F_{\nu}^{J}\!\!=\!m \log\!\delta\!+\!\log \!\left(\!F_{\nu_{p}^{\prime}}^{\prime}\!\right)\!-\!\log\!\left(\!\frac{\nu}{\nu_{p}}\!\right)\!-\!b\left[\!\log\!\left(\!\frac{\nu}{\nu_{p}}\!\right)\!\right]^{2}},
\end{gather}
here, $\nu$ can be taken as the different observing frequencies, including optical, $\gamma$-ray, and X-ray.

\subsubsection{\texorpdfstring{$\gamma$}{}-ray Phonton Index and Optical Spectral Index}\label{subsubsec:SIPI}
\begin{figure*}[h!t]
	\centering 
	\subfigure{
		\includegraphics[width=8cm]{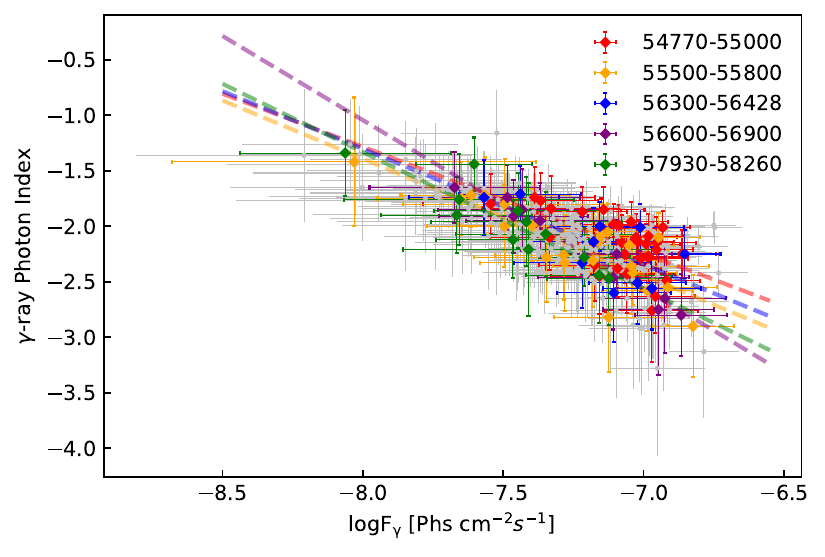}}
	\quad
	\subfigure{
		\includegraphics[width=8cm]{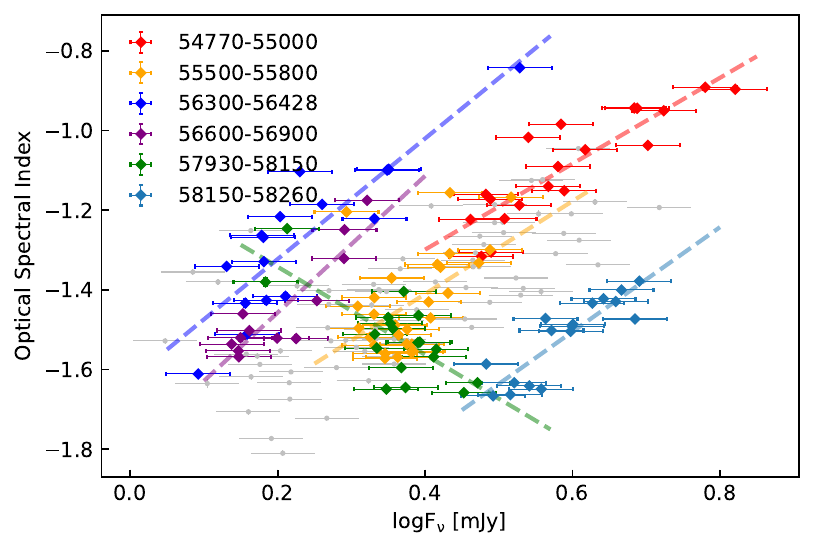}}
	\quad
	\caption{The $\gamma$-ray PI vs. $\log F_{\gamma}$ is plotted on the left panel while the optical spectral index $\alpha_{o}$ vs. $\log F_{\nu}$ is plotted on the right panel. Different colors correspond to different periods.}
	\label{SI}
\end{figure*}

On the left panel of Figure \ref{SI}, the $\gamma$-ray PI versus $\log F_{\gamma}$ is plotted. It is obvious that PI decreases as $F_{\gamma}$ increases, showing a softer when brighter (hereafter SWB) trend. The one-zone emission model is not suitable for producing the SWB trend, therefore, a two-component model could be a viable alternative. Similar to the redder when brighter (hereafter RWB) trend of the optical variation for FSRQs, it can be understood that the increase of the flaring component attenuates the observed spectral index if there is a stable hard background component. The origin of this background component could be studied by very-high-energy $\gamma$-ray \citep{2020A&A...640A.132M}. Due to the fact that LAT instrument response is biased toward hard photon indexes at low fluxes, there is also a possibility that the SWB behavior is caused by this.

\begin{figure}[t!]
	\centering
	\includegraphics[width=1\columnwidth]{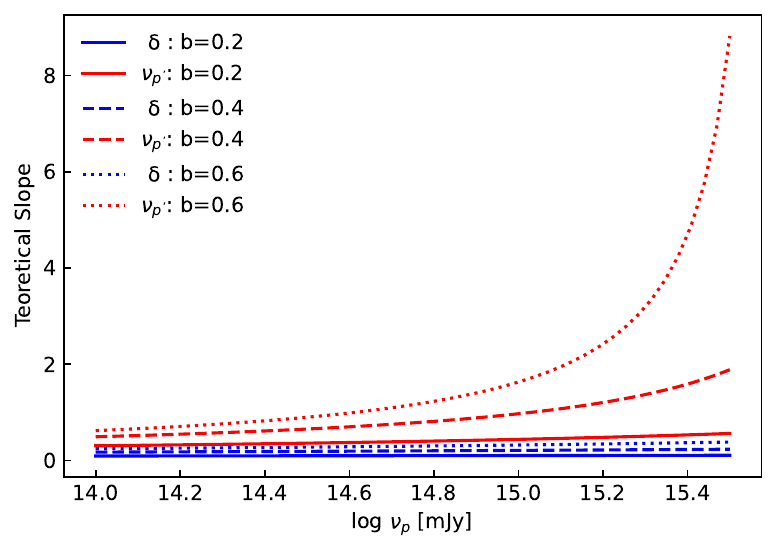}
	\caption{The variation of the theoretical slope of $\alpha_{o} $ versus $\log F_{\nu}$ with $\log \nu_{p}$ in different cases are plotted. The red color represents cases where $\nu_{p}^{\prime}$ dominates the variation and the blue color represents cases where $\delta$ dominates the variation. Solid, dashed, and dotted lines correspond to the cases $b=0.2$, $b=0.4$, and $b=0.6$ respectively.}
\label{slopemodel}
\end{figure}

Additionally, we pair the optical spectral index $\alpha_{o} $ with $\log F_{\nu}$ and plot them on the right panel of Figure \ref{SI}. In contrast to the SWB of the $\gamma$-ray, the optical variation exhibits a harder when brighter (hereafter HWB) trend. These variation behaviors are studied by a linear fitting based on the least-squares method and the results are shown in Table \ref{SIResult}. For period $(E)$, it is divided into two sub-periods $(E1)$ and $(E2)$ when studying the behavior of the optical spectral index because it shows different behaviors in the two sub-periods. Specifically, there is a SWB trend at low luminosity $(E1)$ and a HWB trend at high luminosity $(E2)$. 

\begin{table*}[ht]
	\centering		
	\begin{center}
		\setlength{\abovecaptionskip}{0cm} 
		\setlength{\belowcaptionskip}{-0.2cm}
		\caption{Linear fitting results of PI versus $\log F_{\gamma}$ and $ \alpha_{o} $ versus $\log F_{\nu}$}
		\label{SIResult}
		\setlength\tabcolsep{1mm}{
			\begin{tabular}{c|cccc|cccc}
				\toprule
				\hline
				\multirow{2}{*}{Period} & \multicolumn{4}{c|}{PI versus $\log F_{\gamma}$} & \multicolumn{4}{c}{$ \alpha_{o} $ versus $\log F_{\nu}$}\\
				\cline{2-9}       
				& Slope  & Intercept & Pearson's $r$ & P-value   & Slope  & Intercept & Pearson's $r$ & P-value   \\
				\hline   
				$(A)$54770-55000          &$ -0.957 \pm0.23$                  & $-8.940\pm1.61 $                  & -0.609                        & 2.164e-04                       &$ 1.078\pm0.14$      & $-0.730\pm0.09$        & 0.881         & 6.568e-07    \\ 
				$(B)$55500-55800          & $ -1.051\pm0.15$                    & $-9.804\pm1.10$                       & -0.841                         & 9.755e-07                   & $1.166\pm0.37 $     & $-0.877\pm0.14$     & 0.526         & 4.025e-03    \\
				$(C)$56275-56428          & $-1.040\pm0.38$                    & $-9.627\pm2.69$                       & -0.699                         & 2.458e-07                       & $1.514\pm0.23$      &$ -0.625\pm0.06$        & 0.873         & 1.008e-05    \\
				$(D)$56600-56900          & $-1.517\pm0.11$                    & $-13.178\pm0.82$                      & -0.979                       & 9.014e-07                  & $1.716\pm0.27$      & $-0.800\pm0.06$        &0.896         & 8.144e-05    \\
				$(E1)$57930-58150         & \multirow{2}{*}{$-1.231\pm0.18$} & \multirow{2}{*}{$-11.176\pm1.3$} & \multirow{2}{*}{-0.902} & \multirow{2}{*}{6.122e-05}            & $-1.103\pm0.25 $    & $-0.121\pm0.09$      & -0.543     & 4.873e-04       \\
				$(E2)$58150-58260         &    &         &          &        & $1.312\pm0.19  $    &$ -1.297\pm0.11 $       & 0.873         & 4.678e-06  \\
				\hline
		\end{tabular}}
	\end{center}
\end{table*}

\begin{figure}
	\centering
	\includegraphics[width=1\columnwidth]{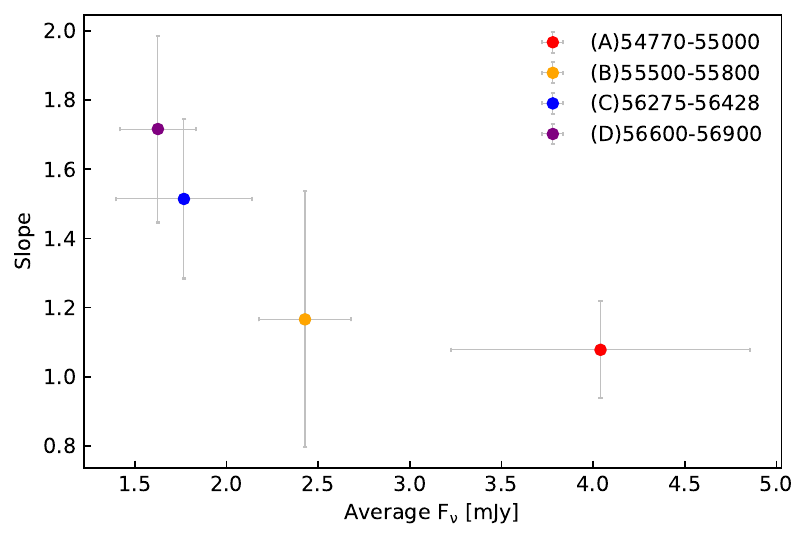}
	\caption{The slopes of $ \alpha_{o} $ vs. $\log F_{\nu}$ (see Table \ref{SIResult}) against the average flux on the corresponding period are plotted. Different colors represent different periods, consistent with those in Figure \ref{SI}. In this panel, period $(E)$ is eliminated due to its V-shaped behavior.}
\label{SI_Slope}
\end{figure}

In addition, we can figure out whether the dominant factor of the variation is extrinsic (the Doppler factor $\delta$) or intrinsic (the intrinsic peak frequency $\nu_{p}^{\prime}$) by the slope in the $\alpha_{o} $ versus $\log F_{\nu}$ plane. We consider that the optical emission is dominated by the synchrotron process in the jet and the flux can be described by Equation \ref{eq:LP}. By definition, the optical spectral index of the jet component $\alpha_{\nu}^{J} $ can be derived as
\begin{gather}\label{eq:SI} 
	{ \alpha_{\nu}^{J}= \frac{d\log F_{\nu}^{J}}{d\log \nu} =-1-2b\log \left(\frac{\nu}{\delta \nu_{p}^{\prime}}\right)},
\end{gather}
here, $\alpha_{\nu}^{J}$ is a function of the curvature parameter $b$, the Doppler factor $\delta$, and the intrinsic peak frequency $\nu_{p}^{\prime}$. To further simplify, we take an empirical relation $F_{\nu_{p}^{\prime}}^{\prime}\propto \nu_{p}^{\prime n}$, where $n$ is usually taken as 0 \citep{Jiang2023}. The function of $\log F_{\nu}^{J}$ relying on $\alpha_{\nu}^{J}$ can be derived as
\begin{gather}\label{eq:FVSI}
	{\!\log\!F_{\nu}^{J}\!\!=\!m\!\log\!\delta\!+\!n\! \log\!\nu_{p}^{\prime}\!+\!\frac{1}{2b}\!\left(\! \alpha_{\nu}^{J}\!+\!1\!\right)\!-\!\frac{1}{4b}\!{\left(\!\alpha_{\nu}^{J}\!+\!1\!\right)\!}^{2}}\!\!+\!C\!,
\end{gather}
here, $C$ is a constant independent of $\nu_{p}$. In the case of the variation dominated by the Doppler factor $\delta$, the slope is derived as $K_{\delta}=\left[(m+1)/2b+\log \left(\nu/\delta \nu_{p}^{\prime}\right)\right ]^{-1}$. The slope is derived as $K_{\nu_{p}^{\prime}}=\left[(n+1)/2b+\log \left(\nu/\delta \nu_{p}^{\prime}\right)\right]^{-1}$ when the variation is modulated by $\nu_{p}^{\prime}$. Since ON 231 is an intermediate synchrotron peaked blazar, the synchrotron peak frequency ($\delta \nu_{p}^{\prime}$) is near the observed  frequency $\nu=10^{14.78}$ Hz (5100 Å). We can ignore the part $\log (\nu/\delta \nu_{p}^{\prime})\sim 0$ and make the slope formula simpler. With $m = 3$ and $n = 0$, we have $K_{\delta}\approx b/2$ and $K_{\nu_{p}^{\prime}}\approx 2b$. Based on the above scenario, if the jet component dominates the variation, the whole process will exhibit a HWB trend whether the variation is modulated by $\delta$ or $\nu_{p}^{\prime}$. The only difference is that the process dominated by  $\nu_{p}^{\prime}$ will show a steeper HWB trend compared to that dominated by $\delta$.

The fitting slopes of $\alpha_{o} $ versus $\log F_{\nu}$ for each period shown in Table \ref{SIResult} all exceed 1 except period $(E1)$. We also calculate the overall fitting slope for the entire period with $K_{obs}=0.57\pm0.08$. 
In Figure \ref{slopemodel}, we exhibit the variation of the theoretical slope of $\alpha_{o} $ versus $\log F_{\nu}$ with $\log \nu_{p}$ when the variation is modulated by $\nu_{p}^{\prime}$ and $\delta$. We also consider different scenarios when $b$ is taken as 0.2, 0.4, and 0.6 because the curvature parameter $b$ usually takes different values in previous studies \citep{2019ApJ...880...19K,2020ApJ...898...48A}. Based on the cases in Figure \ref{slopemodel}, we can see that the theoretical slope $K_{\nu_{p^{\prime}}}$ vary from $0.30$ to $8.82$, and $K_{\delta}$ vary from $ 0.10$ to $0.38$. It is evident that the theoretical slope will match the observed slope better if the variation is modulated by $\nu_{p}^{\prime}$ rather than $\delta$. It indicates that this variation mechanism tends to be explained by the shock-in-jet model, i.e., the relativistic electrons are accelerated by the shock in the jet, causing the intrinsic synchrotron peak frequency $\nu_{p}^{\prime}$ to shift to higher energy. On the right panel of Figure \ref{SI}, the fitted lines on different periods seem to shift horizontally relative to each other. This can be explained by that the optical emission is not only from the flaring component in the jet but also includes a slightly varying background. As shown in the third panel of Figure \ref{lc}, short-term flares are superimposed on the long-term variable component. Such a component plays the role of the constant background when we study short-term optical spectral behaviors. Similarly, this phenomenon that the slope at long-term timescale is lower than that on each period can be the stacking effect of the event on each period. Thus, the long-term behavior may be not fundamental to reveal the variation mechanism of this target. For the V-shaped behavior in period $(E)$, it can be interpreted that the background component dominates the emission from the jet and causes a trend of SWB before MJD 58150, and then the flaring component becomes dominant and shows the HWB trend. In Figure \ref{SI_Slope}, we exhibit the slopes of $ \alpha_{o} $ versus $\log F_{\nu}$ (see Table \ref{SIResult}) against the average flux on the corresponding period. Among these, period $(E)$ is eliminated due to its V-shaped behavior. We notice that the slope decreases as the source becomes brighter. This also implies that the curvature parameter is not a fixed value in different activity states. The specific explanation of this phenomenon will be discussed in Section \ref{CI} in relation to the behavior of CI.

\subsubsection{Color Index}\label{CI}
\begin{figure}
	\centering
	\includegraphics[width=1\columnwidth]{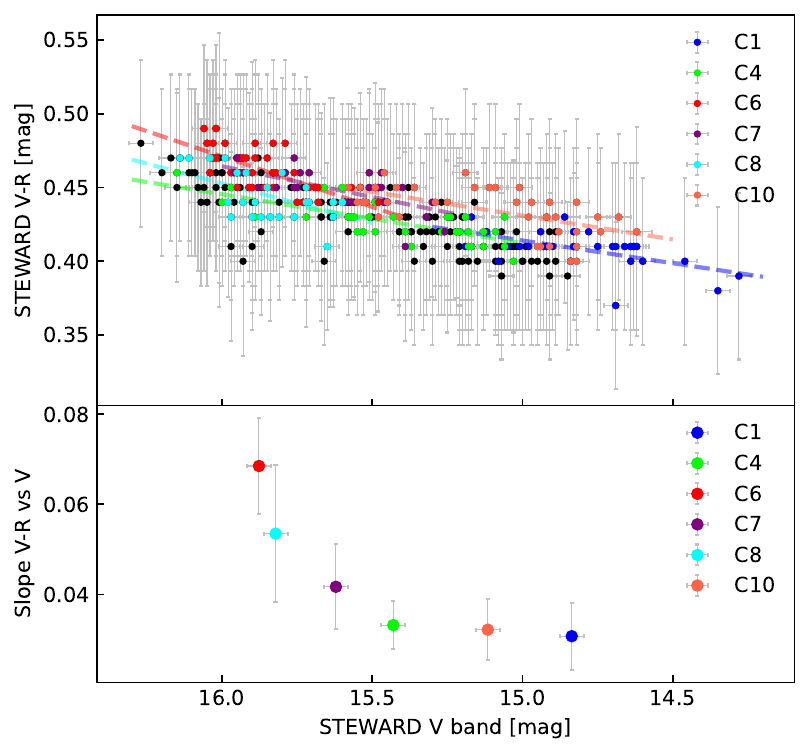}
	\caption{The CI $V-R$ vs. $V$ magnitude is plotted on the top panel and the slope vs. average magnitude on each period is plotted on the bottom panel. Different colors correspond to different periods.}
\label{VR}
\end{figure}
Besides the spectroscopy data, we also consider the photometric data to investigate and confirm the optical variation behaviors, since the photometric data has better sampling. Many works have investigated the color variation behaviors of ON 231, which exhibit a BWB trend \citep{2008PASJ...60..145Z,2013MNRAS.429.2773C,2023MNRAS.519.5263Z}. We pair the magnitudes in $V$-band and $R$-band with the time bin less than 1 day and calculate the CI $V-R$. According to optical data sampling of the SO, the optical light curve can be divided into ten cycles, seeing on the fourth panel of Figure \ref{lc}. We select six cycles (C1, C4, C6, C7, C8, C10) for presenting the CI $V-R$ versus $V$ magnitude on the top panel of Figure \ref{VR}. For each cycle, we calculate the slope of linear fitting and the average magnitude, and plot them on the bottom panel of Figure \ref{VR}. It can be noticed that ON 231 exhibits a BWB trend at both middle-term (one cycle) and long-term timescales. The slope of CI versus magnitude becomes smaller when the source becomes brighter (see the bottom panel of Figure \ref{VR}). This phenomenon is quite similar to the behavior in Figure \ref{SI_Slope}. The decreasing slope with increasing flux indicates that there is a negative correlation between the peak frequency $\nu_{p}$ and spectral curvature $b$, i.e., as more relativistic electrons are accelerated to the high-energy regine by the shock in the jet, the electron spectrum becomes harder and the curvature parameter $b$ becomes smaller. This $\nu_{p}$-b trend can be explained in the framework of energy-dependent acceleration mechanisms where the acceleration probability decreases with the particle energy \citep{2004A&A...413..489M} or through a stochastic acceleration process of relativistic particles undergoing momentum diffusion \citep{2011ApJ...739...66T}. According to the equation $ \alpha_{V-R} = (V-R)/[2.5\log(\nu_{V}/\nu_{R})] $, we translate the slope of CI versus magnitude into the slope of $ \alpha_{V-R}$ versus $\log F_{\nu}$. The slope of the whole period is 0.521 which closely matches the $K_{obs}=0.57\pm0.08$ in Section \ref{subsubsec:SIPI}. The resulted slopes of the six cycles range from 0.399 to 0.889, which are smaller than the slopes in Table \ref{SIResult}. The spectral index of the photometric data measures flux integrated on a range of wavelengths, and it is flatter than that of the spectroscopy data. If we consider that the variation is modulated by $\nu_{p}^{\prime}$ ($K_{\nu_{p}^{\prime}}\in[0.30,8.82]$), the resulted slopes can still be understood in the framework of the shock-in-jet model.

\subsubsection{X-ray Photon Index and Hardness Ratio }\label{HR}
\begin{figure}
	\centering
	\includegraphics[width=1\columnwidth]{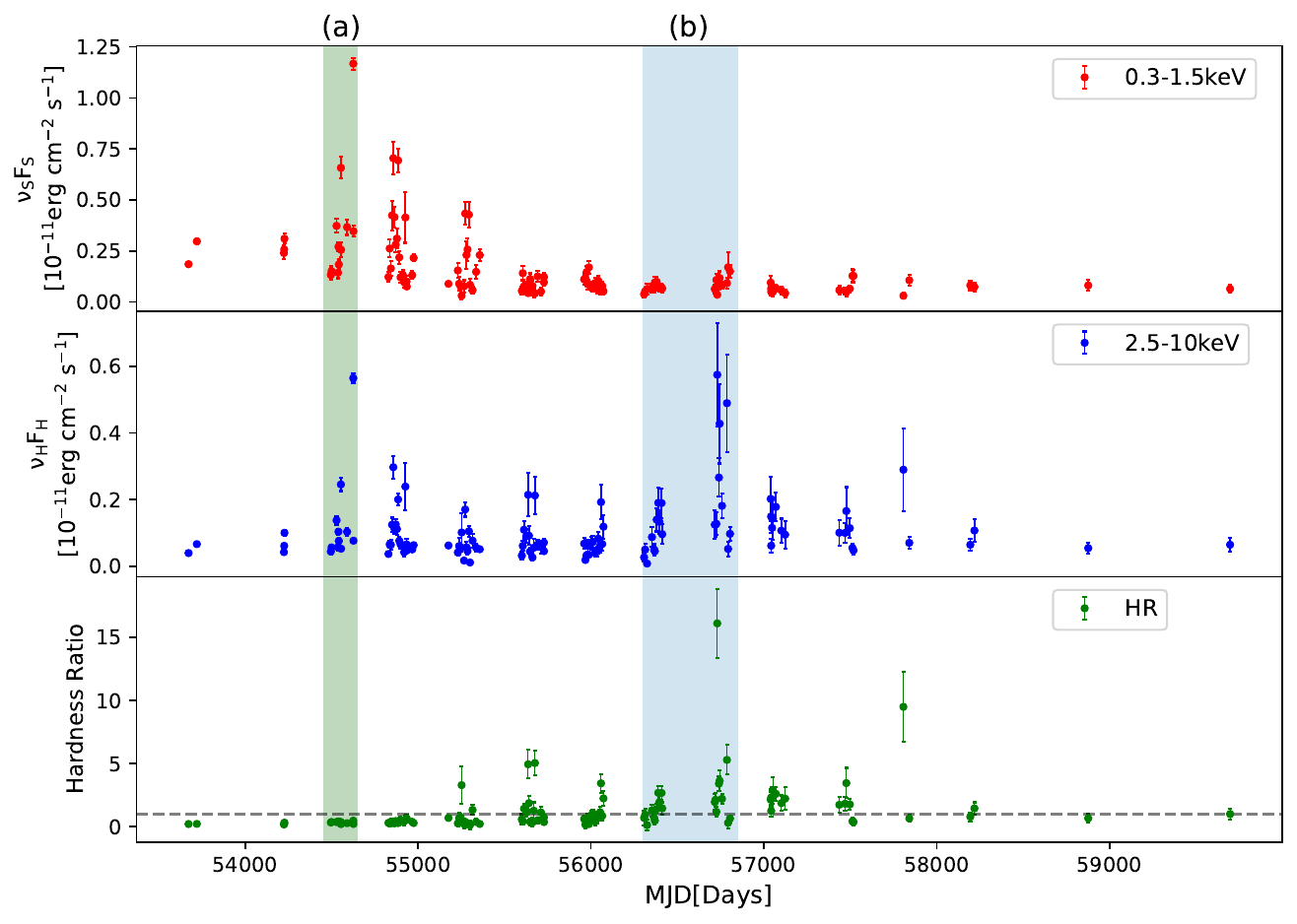}
	\caption{From top to bottom, panels exhibit the light curves of $F_{X}$ in the energy ranges of $0.3-1.5$ keV, $2.5-10.0$ keV, and the HR. Different colors represent different periods. The dashed line in the bottom panel represents HR=1.}
    \label{XrayLC}
\end{figure}

\begin{figure*}[t!]
	\centering
	\includegraphics[width=1.8\columnwidth]{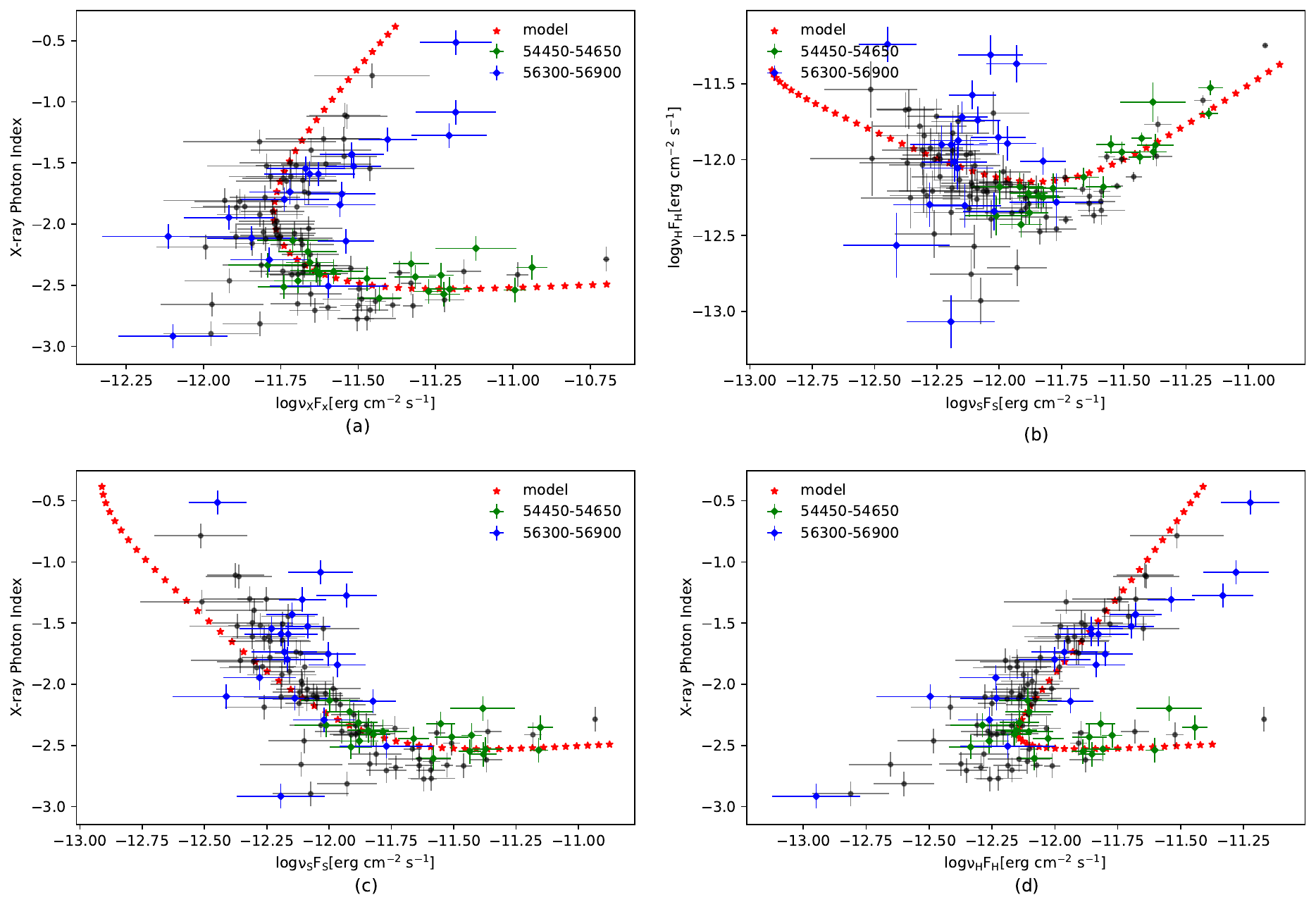}
	\caption{The upper left panel exhibits the behaviors of X-ray PI with flux in $0.3-10$ keV. The upper right panel exhibits the correlation between $\log \nu_{H} F_{H}$ vs. $\log \nu_{S} F_{S}$. The bottom left panel exhibits the behaviors of X-ray PI with flux in $0.3-1.5$ keV while the bottom right panel exhibits the behaviors of X-ray PI with flux in $2.5-10$ keV. Different colors represent different periods. The red asterisked lines represent the numerical results of the model. On all panels, we set the parameters as $b_{1}=0.08, b_{2}=0.55, F_{\nu_{p_{1}}^{\prime}}^{\prime}=10^{-28}$ mJy, $ F_{\nu_{p_{2}}^{\prime}}^{\prime}=10^{-32.8}$ mJy. $\log \nu_{p}^{\prime}$ [Hz] ranges from 12.0 to 13.5.}
	\label{FigXrayPI}
\end{figure*}

The X-ray PI and HR are effective indicators for manifesting the spectral variability of the source along with its brightness. In Figure \ref{XrayLC}, we plot the light curve of X-ray in the soft band, hard band, and HR. We select two typical periods $(a)$, and $(b)$, which correspond to a active state from MJD 54450 to 54650, and a quiescent state but with a flare in hard band from MJD 56300 to 56900. In Figure \ref{FigXrayPI}, the upper left panel exhibits the behaviors of X-ray PI versus the flux in 0.3-10 keV while the upper right panel exhibits the correlation between flux in soft band versus flux in hard band. The behaviors of X-ray PI with the soft band flux and the hard band flux are plotted on the two bottom panels. We notice that the variation behavior on each panel has a turning point and two branches. The variation in period $(a)$ and in period $(b)$ coincide on different branches.
Combined with the bottom panel of Figure \ref{XrayLC}, the HR \textless $1$ in period $(a)$ indicates that the variation is more significant at the soft band. Conversely, the HR \textgreater $1$ in period $(b)$ indicates the variation is more significant at the hard band. The different behaviors during period $(a)$ and period $(b)$ demonstrate that the X-ray radiation in different energy bands could originate from different processes.
\begin{figure*}[t]
	\centering
	\includegraphics[width=1.8\columnwidth]{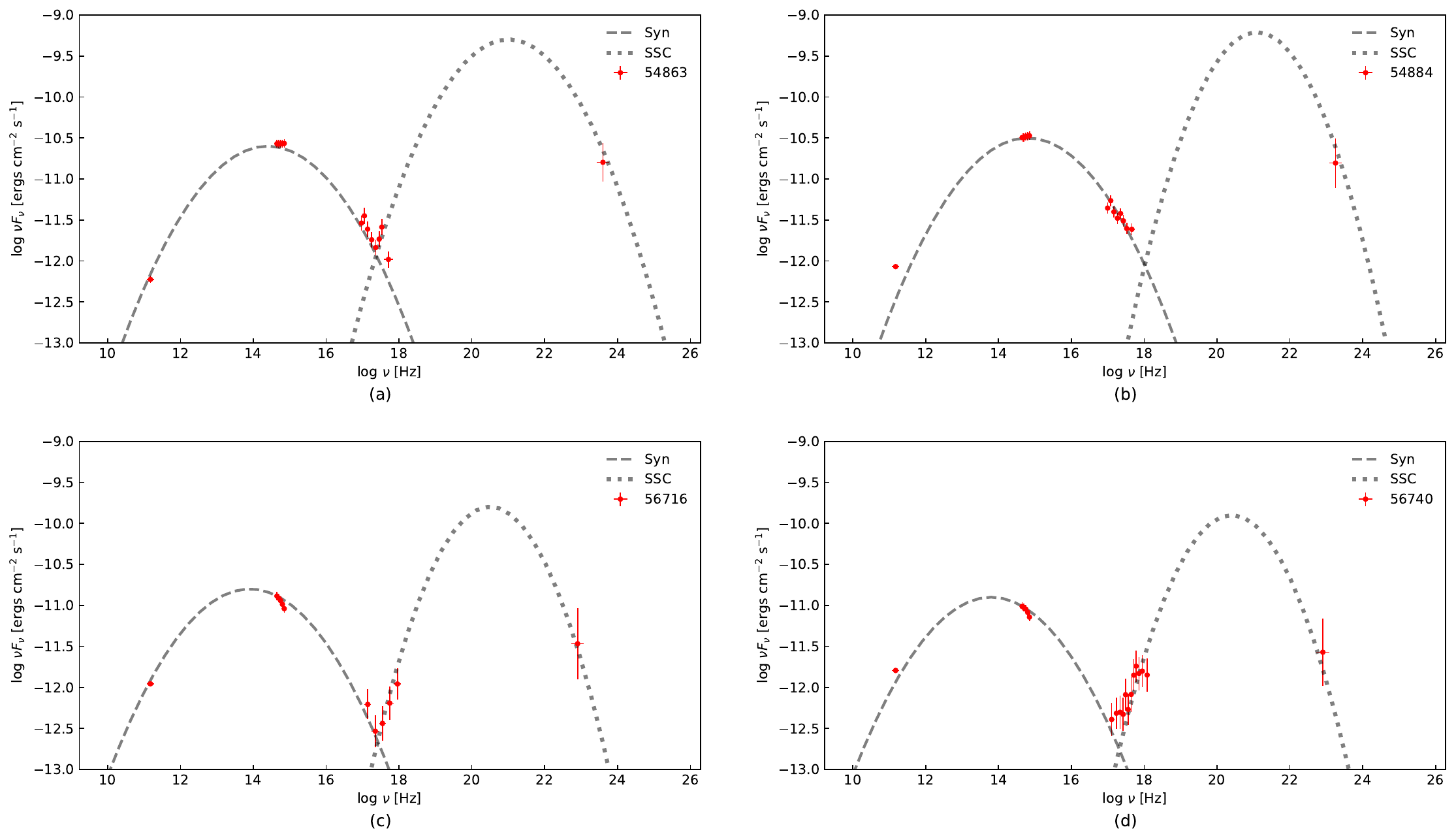}
	\caption{The broadband SED fitted by two log-parabolic models at four different epochs. The dashed line represents the synchrotron component and the dotted line represents the SSC component.}
	\label{SED}
\end{figure*}


Based on the above analysis, we consider that the hard or soft X-ray emission includes both the synchrotron and Synchrotron Self-Compton (hereafter SSC) components, i.e., $F^{obs}_{\nu_{h,s}}=F_{\nu_{h,s}}^{syn}+F_{\nu_{h,s}}^{ssc}$. Previous studies show the fact that the log-parabolic model has advantages in modeling the spectral distributions of the synchrotron and the SSC components \citep{2006A&A...448..861M}. Therefore, we adopt two log-parabolic functions to represent the synchrotron component and the SSC component, respectively. The specific formulas are given as follows
\begin{gather}\label{eq:LP1}
\resizebox{0.89\hsize}{!}{$\log F_{\nu_{h,s}}^{syn}=m \log \delta+\log \left(F_{\nu_{p_{1}}^{\prime}}^{\prime}\right)-\log \left(\frac{\nu_{h,s}}{\nu_{p_{1}}}\right)-b_{1}\left[\log \left(\frac{\nu_{h,s}}{\nu_{p_{1}}}\right)\right]^{2}$},\\
\label{eq:LP2}
\resizebox{0.89\hsize}{!}{$\log F_{\nu_{h,s}}^{ssc}=m \log \delta+\log \left(F_{\nu_{p_{2}}^{\prime}}^{\prime}\right)-\log \left(\frac{\nu_{h,s}}{\nu_{p_{2}}}\right)-b_{2}\left[\log \left(\frac{\nu_{h,s}}{\nu_{p_{2}}}\right)\right]^{2}$},
\end{gather}
here, the $\nu_{p1}, F_{\nu_{p_{1}}^{\prime}}^{\prime}, b_{1}$ represent the observed peak frequency, the intrinsic flux at the peak frequency, and the curvature parameter of the synchrotron component, respectively. The $\nu_{p2}, F_{\nu_{p_{2}}^{\prime}}^{\prime}, b_{2}$ represent those of the SSC component. The observed peak frequency of the synchrotron component and SSC component can be related by $\nu_{p_{2}}/\nu_{p_{1}}=4\gamma^2$, where $\gamma$ is the Lorentz factor of electrons.

\begin{table*}[t!]
	\centering		
	\begin{center}
		\setlength{\abovecaptionskip}{0cm} 
		\setlength{\belowcaptionskip}{-0.2cm}
		\caption{The fitting results of the broadband SED in different epochs}
  \label{SEDTable}
  \setlength\tabcolsep{6.6mm}{
   \begin{tabular}{c|cccccccc}		
    \toprule	
    \hline
    Period & $m$ &$\delta$& $\gamma$ & $b_{1}$ &$b_{2}$& $F_{\nu_{p_{1}}^{\prime}}^{\prime}$ & $F_{\nu_{p_{2}}^{\prime}}^{\prime}$ & $\log \nu_{p_{1}}^{\prime}$  \\
				\hline
				$(A1)$54863& 3 &10 & 1000 & 0.15 & 0.2 & -28.0 & -33.3 & 13.4  \\
				$(A2)$54884& 3 &10 & 700  & 0.15 & 0.3 & -28.3 & -33.3 & 13.8  \\
				$(D1)$56716& 3 &10 & 1000 & 0.15 & 0.3 & -27.7 & -33.3 & 12.9  \\
                $(D2)$56740& 3 &10 & 1000 & 0.15 & 0.3 & -27.7 & -33.3 & 12.8  \\
				\hline
		\end{tabular}}
	\end{center}
\end{table*}

In the framework of the shock-in-jet model, we consider that the variation process is dominated by the variation of the intrinsic peak frequency $\nu_{p}^{\prime}$. In theory, the X-ray PI is calculated by $\alpha_{X}=(\log F_{\nu_{h}}-\log F_{\nu_{s}})/(\log \nu_{h}-\log \nu_{s})$. The hard and soft frequencies corresponding to 5 keV and 0.67 keV (roughly the geometric median of the hard and soft band) are $\log \nu_{h}=18.08$ and $\log \nu_{s}=17.21$, respectively. To model the behaviors of PI versus $\log F_{X}$ and $\log F_{H}$ versus $\log F_{S}$, we first set the parameters as $m=3$, $\delta = 10$, $\gamma=1300$. Then, we manually vary the peak frequency $\nu_{p}^{\prime}$ [Hz] ranging from 12.0 to 13.5, and adjust the other parameters $b_{1}, b_{2}, F_{\nu_{p_{1}}^{\prime}}^{\prime}, F_{\nu_{p_{2}}^{\prime}}^{\prime}$ to match the data as closely as possible. The simulation results, marked by the red asterisked lines, are plotted in Figure \ref{FigXrayPI}. The corresponding free parameter values are shown in the caption of Figure \ref{FigXrayPI}. Considering that the free parameters are coupled, the fitted results still have a certain degree of degeneracy. It is clear that the theoretical model can well reproduce the above variation on all panels of Figure \ref{FigXrayPI}, especially for the turning point and two branches. Taking panel $(a)$ for example, while the synchrotron component dominates the X-ray emisson, the target exhibit a HWB trend with a smaller slope. While the SSC component dominates the X-ray emisson, the target exhibit a HWB trend with a higher slope. If the contribution of the synchrotron component and  SSC component about equal, the target could exhibit a SWB trend. In summary, we tend to conclude that the X-ray emission originates from the superposition of the synchrotron component and the SSC component, and the variation is modulated by the shift of the intrinsic peak-frequency $\nu_{p}^{\prime}$ caused by the shock accelerating in the jet.

\subsubsection{Broadband Spectral Energy Distribution}\label{subsubsec:SED}
To understand the diverse behaviour of the source during different periods, we select two epochs from period $(A)$ (high luminosity state), and the other two epochs from period $(D)$ (low luminosity state). Epochs $(A1)$, $(A2)$, $(D1)$, and $(D2)$ are around MJD 54863, 54884, 56716, and 56740, respectively. The time ranges for these epochs is $1\sim3$ days, and a longer time range is used in rare cases due to the lack of data. In Figure \ref{SED}, we construct SEDs and model them using the theoretical framework mentioned in Section \ref{HR}. Specifically, we use two log-parabolic functions to represent the synchrotron component and SSC component, respectively. The specific formulas are given in Equation \ref{eq:LP1} and Equation \ref{eq:LP2}. The fitting results of the broadband SEDs in different epochs are listed in Table \ref{SEDTable}. The SEDs during these four epochs exhibit different behaviors. In epoch $(A1)$, the optical band is around the synchrotron peak frequency $\nu_{p}$, and the X-ray emission includes both the synchrotron and SSC components. In epoch $(A2)$, the optical band falls on the left side of the synchrotron peak frequency $\nu_{p}$, and the X-ray emission is from the synchrotron tail. In epoch $(D1)$, the optical band is located on the right side of the synchrotron peak frequency $\nu_{p}$, and the X-ray emission includes both the synchrotron and SSC components but the SSC component dominates. In epoch $(D2)$, the optical band falls on the right side of the synchrotron peak frequency $\nu_{p}$, and the X-ray emission is from SSC component. Combined with the fitting results in Table \ref{SEDTable}, we find that the variation of this target is dominanted by the shift of the intrinsic peak-frequency $\nu_{p}^{\prime}$. As $\nu_{p}^{\prime}$ increases, the optical band varies from being on the right side of synchrotron peak frequency $\nu_{p}$ to around $\nu_{p}$, and then to the left side of $\nu_{p}$. Meanwhile, the dominant component contributing to the X-ray emission varies from the SSC component to the synchrotron component. In most cases, the two components are in competition with each other and superimposed to compose the X-ray emission. In conclusion, the broadband SEDs visually demonstrate that the variation is dominated by $\nu_{p}^{\prime}$, and X-ray emission is the superposition of both the synchrotron and SSC components.

\subsection{Polarimetric Variability} \label{subsec:PD}
\subsubsection{Optical Polarization and Optical Flux}\vspace{0em}\label{subsec:PDFV}
The long-term monitoring of optical polarization has shown that polarization behavior is complex, and it serves as a significant window to figure out the jet structure and radiation mechanism in BL Lacs. Generally, the shock-in-jet scenario is favored for the correlation between optical flux and PD, whereas the anticorrelation can be explained by the multi-component model \citep{2022MNRAS.511.5611O}.

\begin{figure}[t]
	\centering
	\includegraphics[width=1\columnwidth]{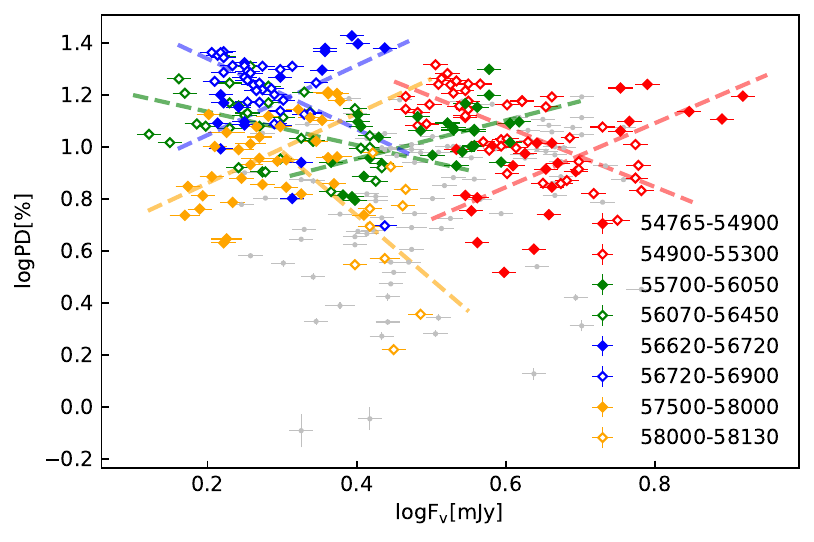}
	\caption{Different colors correspond to different segments. Each segment is divided into two sub-periods. The sub-periods represented by the solid dots exhibit a positive correlation while the sub-periods represented by the hollow dots exhibit a negative correlation.}
    \label{PD_FV}
\end{figure}

We plot $\log$PD versus $ \log F_{\nu}$ ($V$-band flux) in Figure \ref{PD_FV}. The linear fitting results were listed in Table \ref{PDFVslopes}. Four segments are selected and represented by different colors in Figure \ref{PD_FV}, and each segment is divided into two sub-periods. Noteworthy, three segments (marked by the blue, green, and red colors) show the V-shaped behavior, one segment (marked by the yellow color) shows the inverted V-shaped behavior in the correlation between PD and flux, i.e., for each segment, one sub-period exhibits the positive correlation while the other sub-period exhibits the negative correlation. \citet{2014ApJ...780...87M} proposed a Turbulent Extreme Multi-Zone (TEMZ) model to explain the variability of the flux and polarization. The model divides the emitting region into a large number of emission zones (“cells”) with different magnetic field orientations, and the superposition of synchrotron radiation from these cells can lead to a decreasing polarization with increasing flux. In this work, we consider the two-component model. Both components are of jet. One is the polarized background component, representing the post-flaring blobs, and the other component represents a newly appearing flare with different magnetic field orientations. With the increasing luminosity caused by the flaring component, the polarization from the background component is diluted, and the observed PD decreases. When the flaring component becomes dominant, it exhibits a positive correlation between PD and flux, which could be interpreted by the shock-in-jet model.

In Figure \ref{PD_FV}, the three V-shaped behaviors have a translation relation, and it seems that the lower limit of the $\log$PD of the V-shaped behaviors for different segments decreases with increasing flux. A similar translation relation is also found in the plot of $\alpha_{o} $ versus $\log F_{\nu}$ in Figure \ref{SI}. Both the PD and spectral behaviors could be explained if there exists a long-term slightly varying background component. For PD, the V-shaped behavior shifts towards the high flux segment as the background component brightens. For the spectral behaviors, the V-shaped behavior (see Figure \ref{SI}, the green and light blue points, from MJD 57930 to 58260) could be understood if the background component has a bluer spectrum than the new flare. Thus, The translation phenomenon can be understood well by the two-component model, and the variation is dominated by the shock in the jet.
\begin{table}
	\centering		
	\begin{center}
		\setlength{\abovecaptionskip}{0cm} 
		\setlength{\belowcaptionskip}{-0.2cm}
  \caption{Linear fitting results of $\log$PD versus $ \log F_{\nu}$ }
\label{PDFVslopes}
		\setlength\tabcolsep{1.5mm}{
			\begin{tabular}{c|cccc}
				\toprule
				\hline
				Period &Slope  & Intercept &Pearson's $r$ & P-value \\
				\hline    
		54765-54900 & 1.233  & 0.105  &0.689   & 1.961e-04 \\
		54900-55300 & -1.160 & 1.773 & -0.727 & 7.187e-09 \\
		55700-56050 & 0.738  & 0.659 & 0.509  & 3.461e-03 \\
		56070-56450 & -0.643 & 1.263 & -0.423 & 1.278e-02 \\
		56620-56720 & 1.338  & 0.778 & 0.522 &  3.149e-02 \\
		56720-56900 & -1.354 & 1.609 & -0.643 & 7.284e-05 \\
		57500-58000 & 1.336  &  0.593&0.553   & 3.823e-03 \\
        58000-58130 & -2.458 &  1.719 &-0.36   & 3.900e-01\\
				\hline
		\end{tabular}}
	\end{center}

\end{table}

\begin{figure*}[t]
	\centering 
    \subfigure[$(a)$]{
	  \includegraphics[width=5.5cm]{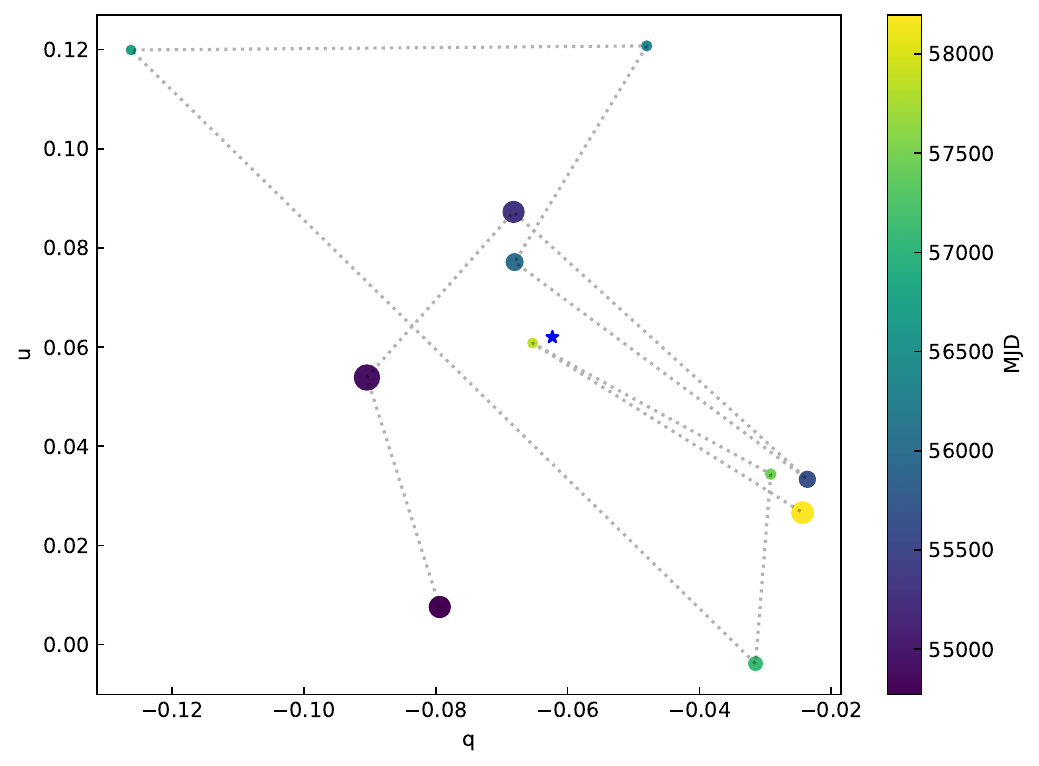}}
	\quad
	\subfigure[$(b)$]{
		\includegraphics[width=5.5cm]{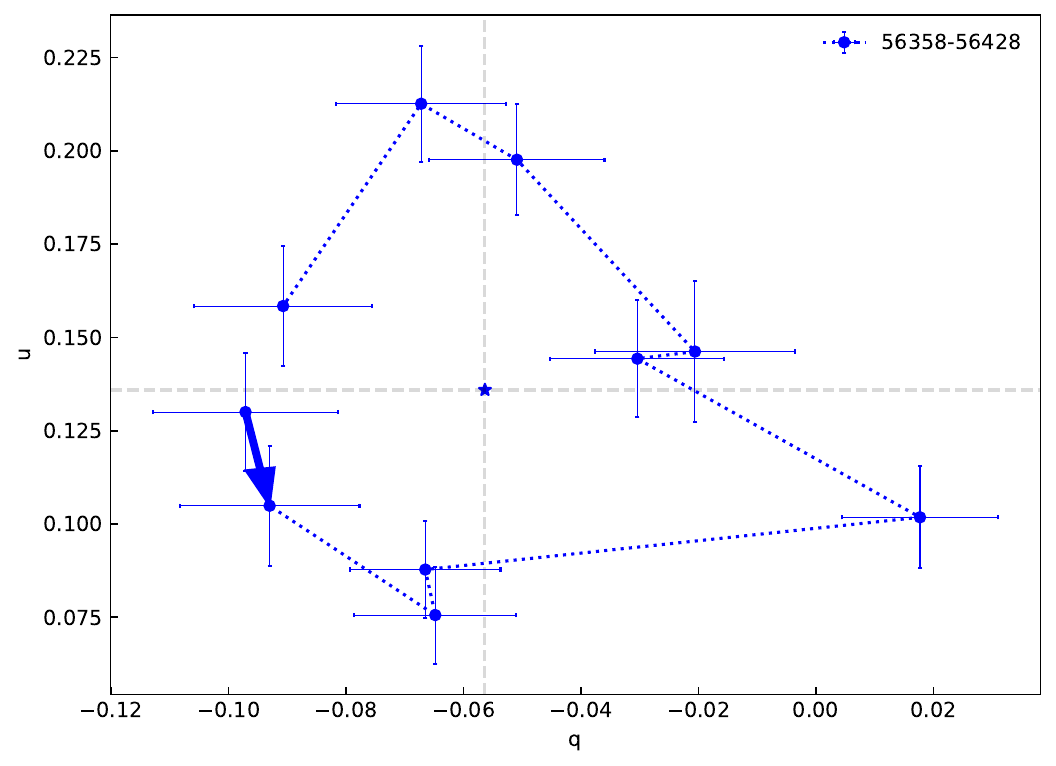}}
	\quad
	\subfigure[$(c)$]{
		\includegraphics[width=5.5cm]{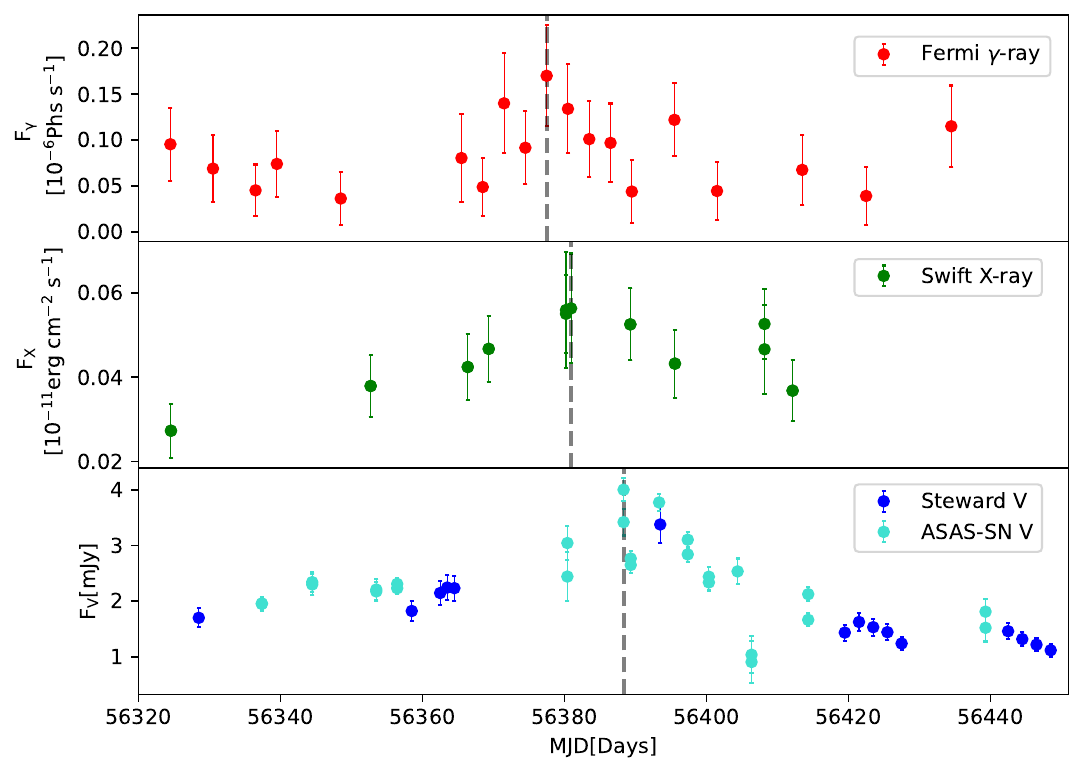}}
	\quad
	\caption{The rotation behavior on $qu$-plane based on one-year binned $qu$ data is shown on the left panel. The circle size represents the averaged flux for the corresponding period. The colors of the circles represent the observation date, where values are listed in the color bar. The middle panel exhibits a typical ACW rotation behavior while the corresponding multiwavelength flares are plotted on the right panel. The blue asterisk represents the center of each rotation behavior.}
	\label{qu}
\end{figure*}

\begin{figure*}[t]
\centering 
\subfigure[$(a)$]{
\includegraphics[width=5.5cm]{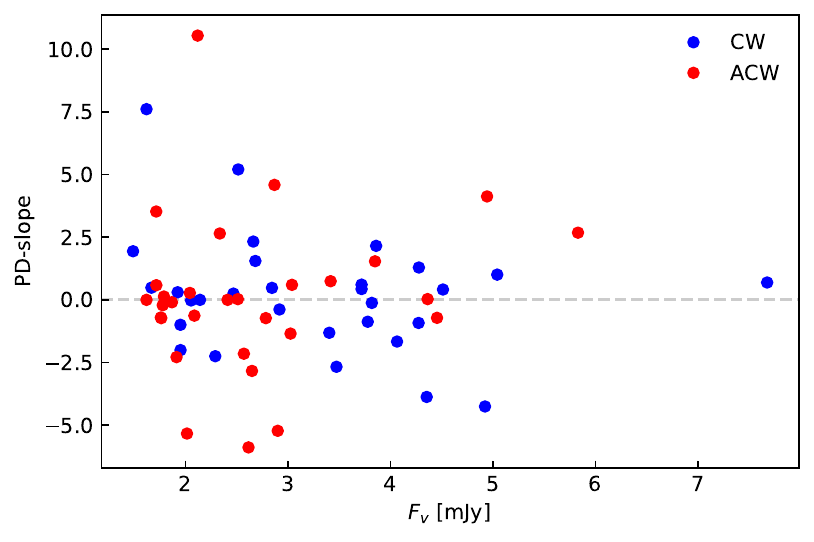}}
\quad
\subfigure[$(b)$]{
\includegraphics[width=5.5cm]{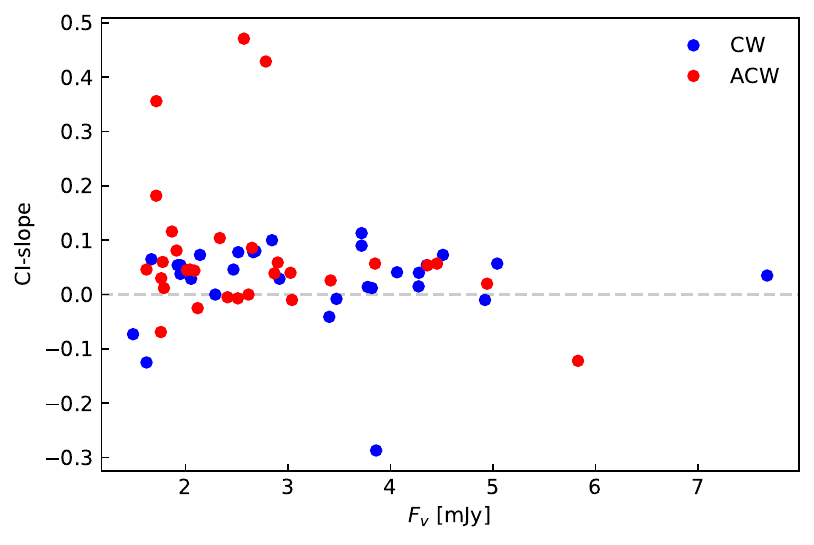}}
\quad
\subfigure[$(c)$]{
\includegraphics[width=5.5cm]{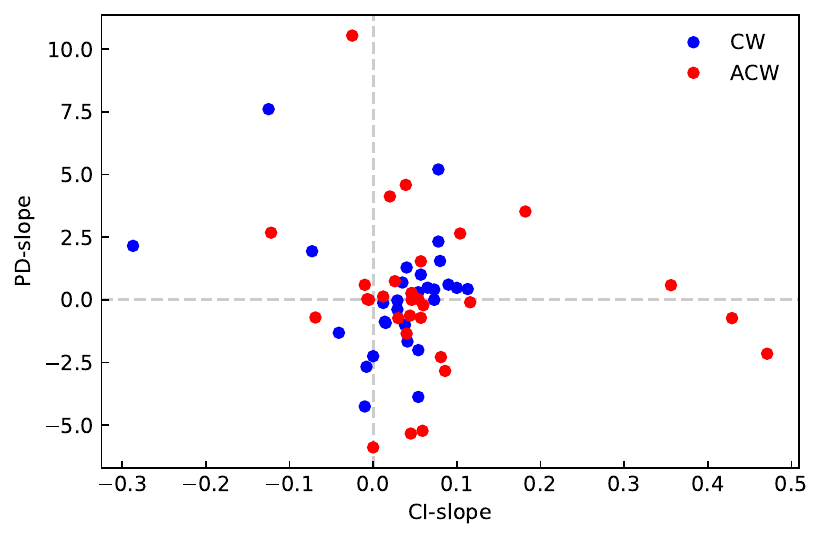}}
\quad

\subfigure[$(d)$]{
\includegraphics[width=5.5cm]{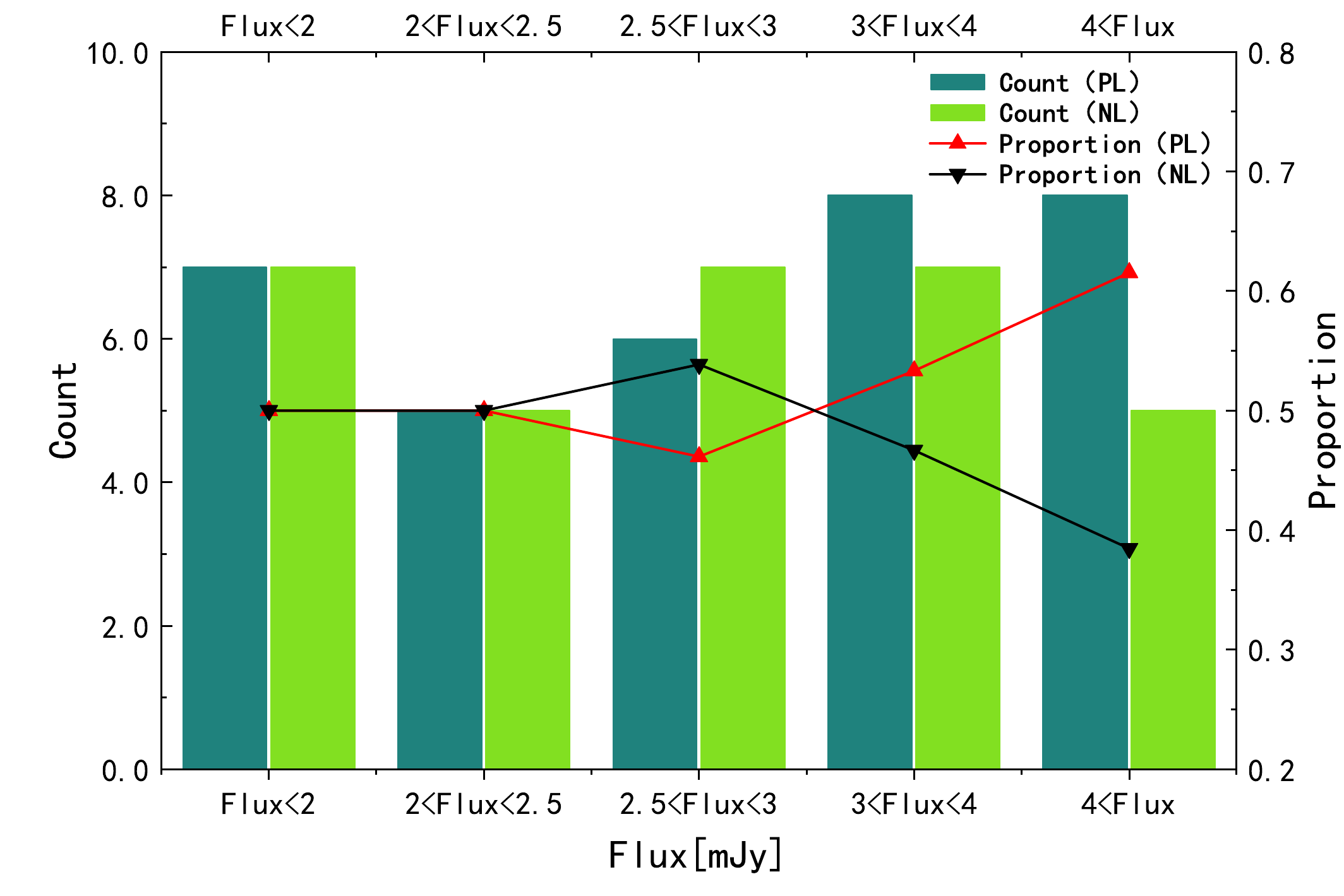}}
\quad
\subfigure[$(e)$]{
\includegraphics[width=5.5cm]{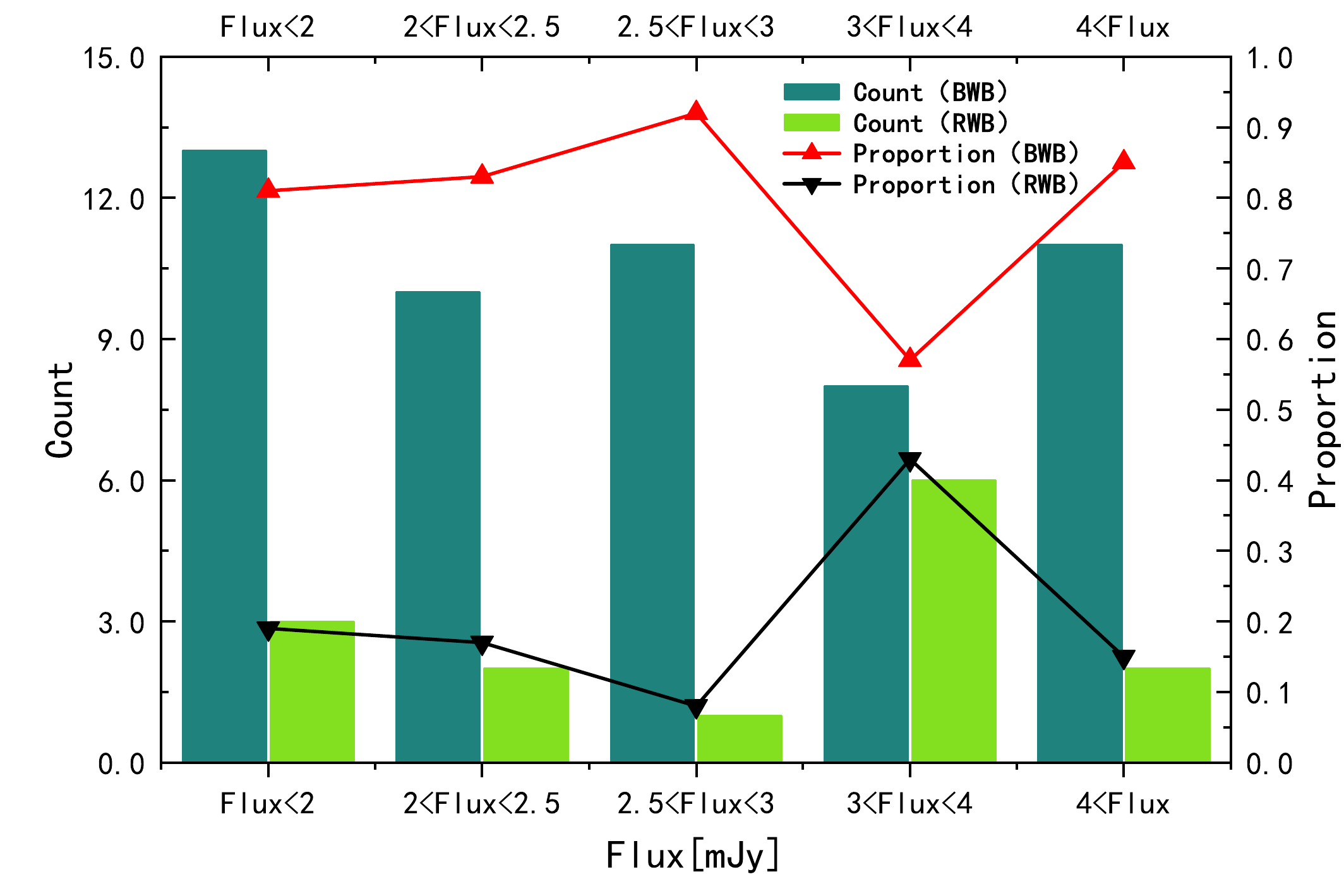}}
\quad
\subfigure[$(f)$]{
\includegraphics[width=5.5cm]{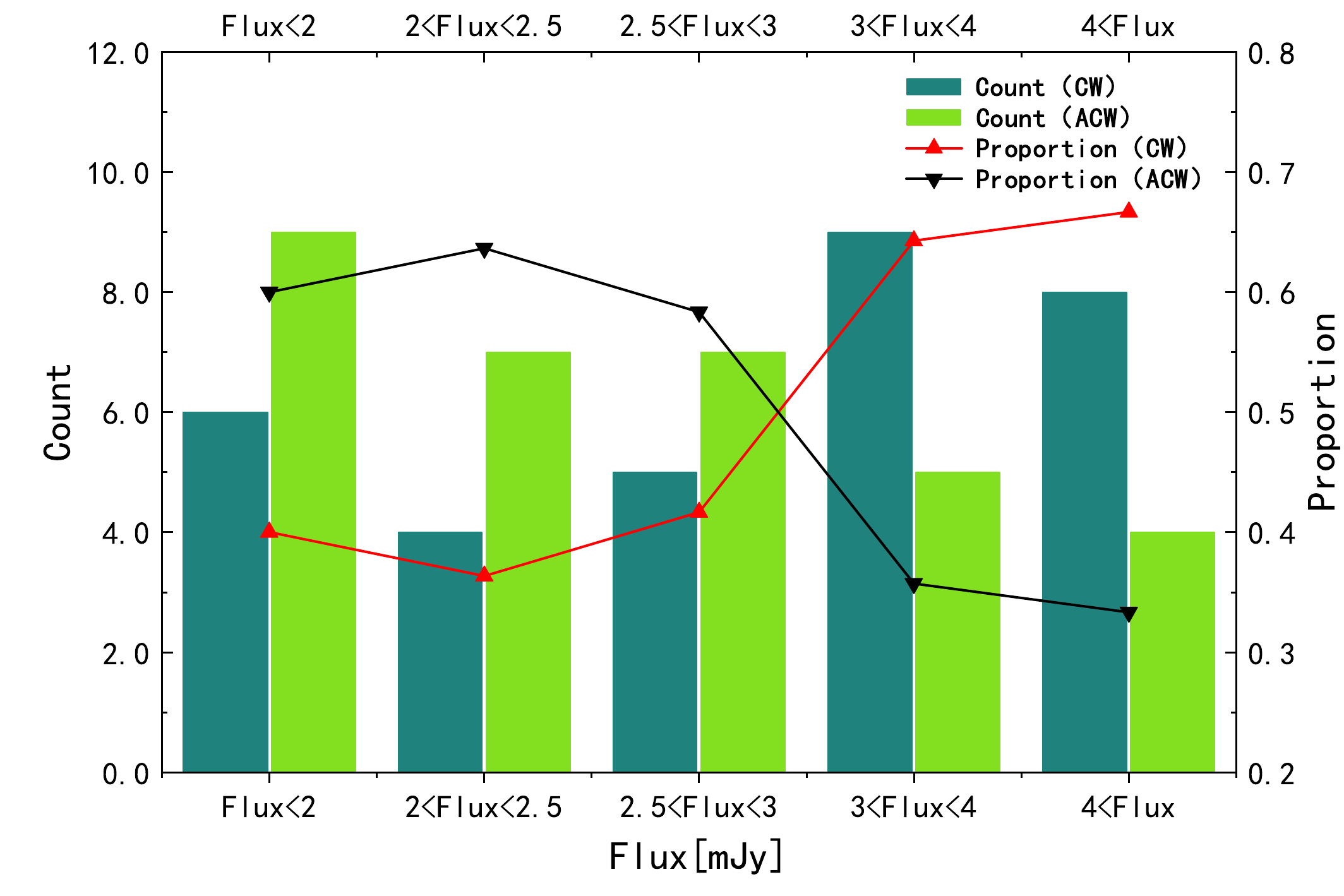}}
\caption{Each dot in the top three panels corresponds to one short period in Table \ref{Tab:Sta}. Panel $(a)$, $(b)$, and $(c)$ exhibit the variation of the PD-slope vs. the average flux, the CI-slope vs. the average flux, and the PD-slope vs. the CI-slope, respectively. The dots with blue color represent the period with the CW rotation behavior while red corresponds to the ACW rotation behavior. The bottom three panels exhibit the statistical results of Table \ref{Tab:Sta}. The bar charts exhibit the number of periods with different behaviors while the line charts exhibit the proportion of periods with different behaviors. Panel $(d)$, $(e)$, and $(f)$ exhibit the variation of the correlation of $\log$PD vs. $\log F_{\nu}$, the CI behavior, and the rotation behavior on $qu$-plane with increasing flux, respectively.}
\label{Fig:Sta}
\end{figure*}

\subsubsection{Rotation Behavior of Polarization Angle}\label{subsec:qu}

The PA, $\chi$, is defined within a $180^{\circ}$-interval, thus the rotation behavior of PA can't be directly presented because of the $n\pi$-ambiguity: $\chi = \chi \pm n\pi, n \in N$ \citep{2008Natur.452..966M,2010Natur.463..919A,2016A&A...590A..10K}. To escape the $n\pi$-ambiguity problem, we study the variation behavior on the $qu$-plane. Here, $q = Q/I$, $u = U/I$, $Q$, $U$, and $I$ are the Stokes parameters. We plot the $q$ versus $u$ for different periods in Figure \ref{qu}. The left panel shows the one-year binned $qu$ data, where the circle size represents the flux value and the colors represent the observation date. The average of $q$ and $u$ of all data points (marked by a blue asterisk) deviates significantly from the origin ($q=u=0$). This translates a constant polarization degree $P_{ave}=10.7\% \pm5.1\%$ and polarization angle $\Theta_{ave}=69.4^{\circ}\pm22.2^{\circ}$.

In the left panel of Figure \ref{qu}, we also find an interesting phenomenon that the rotation of PA is first clockwise at higher luminosity and then anticlockwise or oscillation at lower luminosity. We think that the rotation and oscillation behaviors of PA originate from the flaring activities. Additionally, we select a middle-term timescale period with a typical anticlockwise (hereafter ACW) rotation behavior and plot the $qu$ data on the middle panel of Figure \ref{qu}. Interestingly, the ACW rotation of PA coincides in time with multiwavelength flares. We plot the simultaneous light curves of $\gamma$-ray, X-ray, and optical $V$-band on the right panel of Figure \ref{qu}, among which the $\gamma$-ray peak at MJD 56377.5, the X-ray peak at MJD 56380.9, the optical $V$-band peak at MJD 56388.4. There appear to be time lags between multiwavelength flares, i.e., the $\gamma$-ray band leads the X-ray band while the X-ray band leads the $V$-band. Generally, the above phenomenon suggests that the polarimetric variation is a joint contribution of a stable polarization background component and a variable flaring component. 

\subsection{Statistics of Behavior at Short-term Timescale}\label{subsec:Sta}

To investigate the mechanism of the short-term variation, we divide the light curves into many short-term periods of about one week. The division principle is mainly based on the optical data sampling of the SO. Some periods include a longer time range due to the lack of data points or the guarantee for the integrity of variation. We calculate the parameters (Slope, Pearson’s $r$, P-value) of linear fitting $\log$PD versus $\log F_{\nu}$ and  $V-R$ versus $V$ during every short period. For simplicity, the slope of $\log$PD versus $\log F_{\nu}$ is named as PD-slope, and the slope of $V-R$ versus $V$ is named as CI-slope. These parameters together with the rotation behavior of PA, and the $V$-band flux are listed in Table \ref{tab:S}. In Figure \ref{Fig:Sta}, panel $(a)$ exhibits that about half of the periods show a positive correlation of $\log$PD versus $\log F_{\nu}$ while the other half showed a negative correlation. It seems that the correlation appears to be independent of the rotation behavior of PA. On panel $(b)$, most of the periods show the BWB trend, among which the periods with high CI-slopes show the ACW rotation behavior.  
Additionally, we also notice that the CI-slope becomes flatter as the flux increases, and this phenomenon also seems to be present at middle-term timescales in Figure \ref{VR}. It can be explained by the shock-in-jet model with a varying curvature parameter, as discussed in Section \ref{CI}.
For panel $(c)$, there seems to be a weak positive correlation between PD-slope and CI-slope if those periods with high PD-slope or CI-slope are excluded.

On the three bottom panels of Figure \ref{Fig:Sta}, we sort out the number and proportion of the periods with different behaviors. The bar charts represent the number of different behaviors while the line charts represent the percentage of different behaviors. As readily seen, the periods with the positive correlation of $\log$PD versus $\log F_{\nu}$ and the CW rotation behavior gradually dominate with the increasing flux. However, there are always more periods with BWB trends than that with RWB trends no matter whether in the active or quiet state. According to the statistical results at short-term timescales, the two-component model could also be valid in explaining the transitions of different behaviors with increasing flux.

\section{CONCLUSIONS}\label{sec:conclusion}
In this work, we gather the multiwavelength light curves of ON 231, including $\gamma$-ray, X-ray, optical band, optical PD, optical PA, and 15 GHz radio. We perform the correlation analysis and calculate time lags among them. We discern the variation mechanism by analyzing the spectral behaviors at $\gamma$-ray, X-ray, and optical band. To further constrain the variation mechanism, we investigate the correlation of the optical flux versus PD and the rotation behavior of PA. In addition, we also investigate the variation of the above behaviors at short-term timescales. The principal findings of this work are summarized as follows:              

$\bullet$ According to the cross-correlation analysis results, we find that no correlations are beyond the $3\sigma$ significance. However, there seem to be some $2\sigma$ correlations among the light curves of $\gamma$-ray, $V$-band, and radio with large time lags. Specifically, $\gamma$-ray leads $V$-band ${165.4}^{+56.8}_{\rm -123.3}$ days, $V$-band leads the radio ${246.6}^{+26.4}_{\rm -38.4}$ days, and $\gamma$-ray leads the radio ${387.8}^{+13.7}_{\rm -38.7}$ days. Additionally, there is a negative correlation between PD and optical $V$-band at the near-zero time lag. The above phenomenon implies that the one-zone emission model is not favored.

$\bullet$ The behaviors of the multi-band spectral index are investigated. For the $\gamma$-ray PI, it shows a SWB trend. The variation mechanism could be understood by the two-component model, in which a stable hard background component reserves further study. For the optical spectral index, it usually exhibits a HWB trend. But, in one period, it shows a trend transition from SWB to HWB. Additionally, we find that the slope of $\alpha_{o}$ versus $\log F_{\nu}$ decreases as the flux increases. Combined with the behavior of CI, it shows a BWB trend, and the CI-slope also becomes smaller when the source becomes brighter. We present a theoretical frame to quantitatively explain these spectral behaviors. We find that both the geometric ($\delta$) and intrinsic ($\nu_{p}^{\prime}$) variations can produce a HWB trend. Considering the slope of $\alpha_{o}$ versus $\log F_{\nu}$, the shock-in-jet model is more promising than the helical jet model as the dominant mechanism of variation. The decreasing slope indicates the energy-dependent acceleration processes of the radiative particles. The most complex variation is in the X-ray spectrum, it shows a trend transition from HWB to SWB to HWB. We consider that the X-ray spectrum is the superposition of both the synchrotron and SSC components described by two log-parabolic functions. In the case of the variation modulated by $\nu_{p}^{\prime}$, the theoretical framework could well describe the spectral behavior of X-ray. Additionally, We constructed SEDs in different epochs, and the variation also mainly modulated by $\nu_{p}^{\prime}$. It again confirms that the variation mechanism tends to be explained by the shock-in-jet model.

$\bullet$ We find that the correlation between the optical PD and flux exhibits the typical V-shaped behavior. We also notice that there is a rotation behavior of PA that coincides in time with the multiwavelength flares. In addition, the statistics results of short-term behavior exhibit a similar trend transition when the source becomes brighter. These polarization phenomena could also be explained by the two-component model in the shock-in-jet scenario, in which one is a highly polarized background component, and the other is a significantly varied flaring component.

\section{ACKNOWLEDGEMENTS}
This work has been funded by the National Natural Science Foundation
of China under grant no. U2031102, and by the 
Natural Science Foundation of Shandong Provincial under grant no. ZR2020MA062. Data from the Steward Observatory spectropolarimetric monitoring project are used. This program is supported by Fermi Guest Investigator grants NNX08AW56G, NNX09AU10G, NNX12AO93G, and NNX15AU81G. We gratefully acknowledge the optical observations provided by the ASAS. We also acknowledge the use of public data from the Swift data archive. This paper has made use of data from the OVRO 40 m monitoring program, which is supported in part by NASA grants NNX08AW31G, NNX11A043G, and NNX14AQ89G and NSF grants AST-0808050 and AST-1109911. 

\begin{longrotatetable}
\setlength{\tabcolsep}{1.5mm}{
\begin{deluxetable*}{ccc|c|cccc|cccc|ccc}
        \label{Tab:Sta}
        \renewcommand{\arraystretch}{0.937}
	\rotate
	\tablecaption{
    The statistical results of the behaviors at short-term timescale. Here, PC or NC denotes the positive or negative correlation, respectively.
    \label{tab:S}}
	\tablehead{\multirow{2}{*}{Period}
   & \multirow{2}{*}{MJD} & \multirow{2}{*}{Duration} & \multirow{2}{*}{QU-Rotation}& \multicolumn{4}{c|}{$\log$PD vs $\log F_{\nu} $} &\multicolumn{4}{c|}{$V-R$ vs $V$}& \multicolumn{3}{c}{$V$-band Flux [mJy]}\\
   \cline{5-15}
   &&&&Slope  & Pearson's $r$ & P-value & Correlation &Slope  & Pearson's $r$ & P-value &Tendency&Min & Max & Average}
	\startdata	
	1.1  & 54769-54775 & 6  & CW     & 2.154   & 0.976  & 0.024 & PC   & -0.287 & -0.597 & 0.403 & RWB & 3.511 & 4.221 & 3.861 \\
			1.2  & 54794-54806 & 12 & ACW    & -0.717  & -0.157 & 0.736 & NC   & 0.057  & 0.082  & 0.791 & BWB & 3.814 & 4.936 & 4.455 \\
			1.3  & 54828-54833 & 5  & CW     & -0.122  & -0.029 & 0.962 & NC   & 0.012  & 0.012  & 0.976 & BWB & 3.415 & 4.502 & 3.819 \\
			1.4  & 54859-54864 & 5  & ACW    & 2.680   & 0.504  & 0.496 & PC   & -0.122 & -0.095 & 0.823 & RWB & 5.668 & 6.158 & 5.830 \\
            1.5  & 54881-54889 & 8  & CW     & 0.692   & 0.558  & 0.558 & PC   & 0.035  & 0.031  & 0.948 & BWB & 7.005 & 8.268 & 7.675 \\
			1.6  & 54911-54917 & 6  & CW     & 1.006   & 0.193  & 0.755 & PC   & 0.057  & 0.039  & 0.926 & BWB & 4.802 & 5.363 & 5.042 \\
			1.7  & 54946-54955 & 9  & CW-ACW & 0.218   & 0.071  & 0.894 & PC   & 0.008  & 0.009  & 0.979 & BWB & 4.982 & 6.045 & 5.750 \\
			2.1  & 55125-55131 & 6  & CW-ACW & -1.026  & -0.836 & 0.078 & NC   & -0.034 & -0.027 & 0.945 & RWB & 3.202 & 3.676 & 3.401 \\
			2.2  & 55150-55155 & 5  & ACW    & 0.030   & 0.039  & 0.961 & PC   & 0.054  & 0.570  & 0.430 & BWB & 4.182 & 4.586 & 4.363 \\
			2.3  & 55180-55187 & 7  & CW     & -1.313  & -0.803 & 0.102 & NC   & -0.041 & -0.036 & 0.907 & RWB & 3.232 & 3.543 & 3.404 \\
			2.4  & 55210-55214 & 4  & CW     & -24.075 & -1.000 & 1.000 & NC   & -1.000 & -1.000 & 1.000 & RWB & 3.415 & 3.447 & 3.431 \\
			2.5  & 55240-55247 & 7  & ACW    & -1.346  & -0.419 & 0.408 & NC   & 0.040  & 0.041  & 0.890 & BWB & 2.920 & 3.143 & 3.026 \\
			2.6  & 55270-55277 & 7  & CW     & -0.873  & -0.460 & 0.299 & NC   & 0.014  & 0.011  & 0.970 & BWB & 3.511 & 4.260 & 3.779 \\
			2.7  & 55291-55298 & 7  & CW     & -1.664  & -0.770 & 0.043 & NC   & 0.041  & 0.035  & 0.901 & BWB & 3.479 & 4.714 & 4.065 \\
			2.8  & 55330-55337 & 7  & ACW    & 0.744   & 0.094  & 0.881 & PC   & 0.026  & 0.018  & 0.953 & BWB & 3.292 & 3.609 & 3.417 \\
			2.9  & 55357-55364 & 7  & CW     & 0.412   & 0.197  & 0.673 & PC   & 0.073  & 0.090  & 0.681 & BWB & 4.260 & 4.891 & 4.513 \\
			3.1  & 55510-55516 & 6  & CW     & -0.921  & -0.842 & 0.158 & NC   & 0.015  & 0.111  & 0.794 & BWB & 3.173 & 4.714 & 4.275 \\
			3.2  & 55532-55539 & 7  & ACW    & 2.079   & 0.936  & 0.229 & PC   & 0.116  & 0.995  & 0.065 & BWB & 3.002 & 3.814 & 3.421 \\
			3.3  & 55563-55570 & 7  & ACW    & -5.224  & -0.901 & 0.285 & NC   & 0.059  & 0.988  & 0.099 & BWB & 2.713 & 3.202 & 2.901 \\
			3.4  & 55594-55601 & 7  & ACW-CW & -0.658  & -0.425 & 0.401 & NC   & 0.106  & 0.109  & 0.667 & BWB & 2.591 & 2.974 & 2.778 \\
			3.5  & 55623-55629 & 6  & CW     & 0.478   & 0.147  & 0.853 & PC   & 0.100  & 0.114  & 0.789 & BWB & 2.591 & 3.002 & 2.845 \\
			3.6  & 55649-55660 & 11 & ACW    & -2.835  & -0.954 & 0.012 & NC   & 0.086  & 0.108  & 0.782 & BWB & 2.385 & 3.232 & 2.650 \\
			3.7  & 55707-55713 & 6  & CW     & 5.203   & 0.920  & 0.080 & PC   & 0.078  & 0.102  & 0.810 & BWB & 2.407 & 2.688 & 2.516 \\
			3.8  & 55727-55744 & 17 & ACW    & 0.599   & 0.746  & 0.254 & PC   & -0.010 & -0.198 & 0.802 & RWB & 2.320 & 3.609 & 3.040 \\
   	  4.1  & 55889-55895 & 6  & ACW    & -5.886  & -1.000 & 1.000 & NC   & 0.000  & null   & null  & BWB & 2.520 & 2.713 & 2.616 \\
			4.2  & 55922-55928 & 6  & CW     & 1.551   & 0.514  & 0.486 & PC   & 0.080  & 0.893  & 0.107 & BWB & 2.520 & 3.030 & 2.683 \\
			4.3  & 55948-55955 & 7  & CW     & 22.841  & 1.000  & 1.000 & PC   & 0.000  & null   & null  & BWB & 3.609 & 3.643 & 3.626 \\
			4.4  & 55970-55979 & 9  & ACW    & 1.537   & 0.700  & 0.300 & PC   & 0.057  & 0.515  & 0.485 & BWB & 3.415 & 4.031 & 3.850 \\
			4.5  & 56008-56015 & 7  & CW     & -2.670  & -0.837 & 0.077 & NC   & -0.008 & -0.005 & 0.988 & RWB & 3.322 & 3.609 & 3.474 \\
			4.6  & 56039-56048 & 9  & CW     & 0.431   & 0.160  & 0.798 & PC   & 0.113  & 0.874  & 0.126 & BWB & 3.384 & 4.144 & 3.719 \\
			4.7  & 56060-56076 & 16 & CW     & 0.607   & 0.157  & 0.711 & PC   & 0.090  & 0.757  & 0.030 & BWB & 3.384 & 4.144 & 3.719 \\
			4.8  & 56092-56100 & 8  & CW     & 0.303   & 0.203  & 0.743 & PC   & 0.054  & 0.755  & 0.140 & BWB & 1.696 & 2.195 & 1.925 \\
			5.1  & 56265-56273 & 8  & ACW    & 0.276   & 0.068  & 0.932 & PC   & 0.046  & 0.030  & 0.926 & BWB & 1.894 & 2.195 & 2.044 \\
			5.2  & 56304-56332 & 28 & ACW    & 0.132   & 0.148  & 0.779 & PC   & 0.012  & 0.248  & 0.635 & BWB & 1.575 & 2.002 & 1.791 \\
			5.3  & 56358-56365 & 7  & ACW    & -2.051  & -0.884 & 0.116 & NC   & 0.045  & 0.088  & 0.836 & BWB & 2.135 & 2.615 & 2.465 \\
			5.4  & 56393-56427 & 14 & ACW    & -0.630  & -0.815 & 0.048 & NC   & 0.044  & 0.988  & 0.000 & BWB & 1.450 & 3.994 & 2.088 \\
			5.5  & 56442-56449 & 7  & CW     & 1.939   & 0.891  & 0.109 & PC   & -0.073 & -0.158 & 0.842 & RWB & 1.323 & 1.696 & 1.490 \\
			6.1  & 56622-56630 & 8  & CW     & -0.995  & -0.580 & 0.420 & NC   & 0.038  & 0.149  & 0.725 & BWB & 1.650 & 2.115 & 1.952 \\
			6.2  & 56651-56660 & 9  & ACW    & 0.585   & 0.114  & 0.808 & PC   & 0.356  & 0.309  & 0.262 & BWB & 1.635 & 1.776 & 1.717 \\
			6.3  & 56691-56698 & 7  & ACW    & 2.647   & 0.886  & 0.307 & PC   & 0.104  & 0.996  & 0.058 & BWB & 2.256 & 2.474 & 2.336 \\
			6.4  & 56712-56718 & 6  & ACW    & 0.028   & 0.113  & 0.928 & PC   & -0.007 & -0.007 & 0.988 & RWB & 2.277 & 2.738 & 2.512 \\
			6.5  & 56741-56750 & 9  & ACW    & -0.704  & -0.841 & 0.004 & NC   & -0.069 & -0.083 & 0.719 & RWB & 1.605 & 2.058 & 1.761 \\
			6.6  & 56772-56782 & 10 & ACW    & -0.729  & -0.361 & 0.380 & NC   & 0.030  & 0.041  & 0.863 & BWB & 1.620 & 1.877 & 1.765 \\
			6.7  & 56798-56806 & 8  & CW     & -0.023  & -0.015 & 0.977 & NC   & 0.029  & 0.020  & 0.947 & BWB & 1.860 & 2.215 & 2.056 \\
			6.8  & 56827-56835 & 8  & ACW    & -0.097  & -0.046 & 0.931 & NC   & 0.116  & 0.104  & 0.682 & BWB & 1.792 & 1.983 & 1.869 \\
			7.1  & 56986-57019 & 33 & ACW    & -0.729  & -0.073 & 0.927 & NC   & 0.429  & 0.990  & 0.010 & BWB & 2.663 & 2.974 & 2.785 \\
			7.2  & 57042-57051 & 9  & CW     & 0.249   & 0.213  & 0.787 & PC   & 0.046  & 0.993  & 0.007 & BWB & 2.115 & 3.173 & 2.469 \\
			7.3  & 57065-57073 & 8  & ACW    & 4.586   & 0.574  & 0.312 & PC   & 0.039  & 0.056  & 0.871 & BWB & 2.663 & 3.232 & 2.869 \\
			7.4  & 57102-57109 & 9  & ACW    & -2.284  & -0.657 & 0.343 & NC   & 0.081  & 0.174  & 0.681 & BWB & 1.760 & 2.195 & 1.914 \\
			7.5  & 57125-57132 & 7  & ACW    & -2.148  & -0.193 & 0.807 & NC   & 0.471  & 0.792  & 0.208 & BWB & 2.385 & 2.763 & 2.571 \\
			7.6  & 57160-57167 & 7  & ACW    & -5.334  & -0.857 & 0.345 & NC   & 0.045  & 0.945  & 0.212 & BWB & 1.776 & 2.215 & 2.016 \\
			7.7  & 57187-57196 & 9  & ACW    & 10.542  & 0.357  & 0.556 & PC   & -0.025 & -0.104 & 0.868 & RWB & 2.002 & 2.195 & 2.120 \\
			8.1  & 57364-57370 & 6  & CW     & null    & null   & null  & null & 0.073  & 0.054  & 0.908 & BWB & 2.039 & 2.195 & 2.143 \\
			8.2  & 57399-57406 & 7  & ACW-CW & null    & null   & null  & null & 0.135  & 0.971  & 0.154 & BWB & 1.712 & 1.947 & 1.812 \\
			8.3  & 57426-57435 & 9  & ACW    & null    & null   & null  & null & -0.005 & -0.008 & 0.978 & RWB & 2.215 & 2.615 & 2.412 \\
			8.4  & 57457-57464 & 7  & ACW    & null    & null   & null  & null & 0.046  & 0.904  & 0.096 & BWB & 1.491 & 1.826 & 1.621 \\
			8.5  & 57491-57519 & 28 & CW     & -2.005  & -0.806 & 0.099 & NC   & 0.054  & 0.066  & 0.740 & BWB & 1.727 & 2.341 & 1.953 \\
			9.1  & 57716-57723 & 7  & CW     & 7.607   & 1.000  & 1.000 & PC   & -0.125 & -1.000 & 1.000 & BWB & 1.561 & 1.680 & 1.621 \\
			9.2  & 57749-57786 & 37 & CW     & 0.486   & 0.467  & 0.350 & PC   & 0.065  & 0.792  & 0.061 & BWB & 1.477 & 2.020 & 1.669 \\
			9.3  & 57810-57843 & 33 & ACW    & 3.524   & 0.428  & 0.397 & PC   & 0.182  & 0.370  & 0.293 & BWB & 1.620 & 1.860 & 1.716 \\
			9.4  & 57871-57876 & 5  & CW     & -2.248  & -0.886 & 0.307 & NC   & 0.000  & 0.000  & 1.000 & BWB & 2.215 & 2.363 & 2.292 \\
			9.5  & 57886-57899 & 13 & ACW-CW & 1.454   & 0.765  & 0.027 & PC   & 0.109  & 0.088  & 0.712 & BWB & 1.983 & 2.385 & 2.232 \\
			9.6  & 57915-57955 & 40 & ACW    & -0.213  & -0.106 & 0.866 & NC   & 0.060  & 0.222  & 0.720 & BWB & 1.590 & 1.912 & 1.779 \\
			10.1 & 58072-58080 & 8  & CW     & 2.325   & 0.341  & 0.508 & PC   & 0.078  & 0.065  & 0.796 & BWB & 2.497 & 2.920 & 2.663 \\
			10.2 & 58096-58140 & 44 & CW     & -0.384  & -0.058 & 0.873 & NC   & 0.029  & 0.045  & 0.844 & BWB & 2.591 & 3.384 & 2.918 \\
			10.3 & 58157-58197 & 40 & CW     & -3.873  & -0.797 & 0.203 & NC   & 0.054  & 0.858  & 0.142 & BWB & 3.576 & 5.028 & 4.354 \\
			10.4 & 58214-58222 & 8  & CW     & 1.290   & 0.718  & 0.172 & PC   & 0.040  & 0.092  & 0.766 & BWB & 3.676 & 5.720 & 4.278 \\
			10.5 & 58249-58257 & 8  & ACW    & 4.126   & 0.617  & 0.103 & PC   & 0.020  & 0.031  & 0.886 & BWB & 4.339 & 5.720 & 4.943 \\
			10.6 & 58281-58310 & 29 & CW     & -4.253  & -0.608 & 0.110 & NC   & -0.010 & -0.022 & 0.937 & RWB & 4.461 & 6.045 & 4.923\\
	\enddata

\end{deluxetable*}}

\end{longrotatetable}

\bibliography{sample631}{}
\bibliographystyle{aasjournal}

\end{document}